\documentstyle{mn}

\input{psfig}

\newcommand{\tF}{\tilde{F}}
\newcommand{\tV}{\hat{V}}

\newcommand{\avg}[1]{\left\langle{#1}\right\rangle}

\newcommand{\r}{\mbox{\boldmath $r$}}

\renewcommand{\bar}{\overline }
\newcommand{\bn}{\bar{N}}
\newcommand{\xiav}{\bar{\xi}}
\newcommand{\hmpc}{h^{-1}\,{\rm Mpc}}
\newcommand{\Gen}{P}
\newcommand{\rmd}{{\rm d}}

\begin{document}


\title[Cosmic Statistics of Statistics]{Cosmic Statistics of Statistics\\}

\author[I. Szapudi, S. Colombi and F. Bernardeau]{Istv\'an Szapudi,$^1$
St\'ephane Colombi,$^2$ and Francis Bernardeau$^3$\\
$^1$University of Durham, Department of Physics, South Road,
Durham, DH1 3LE, UK\\
$^2$Institut d'Astrophysique de Paris, CNRS, 98bis bd Arago,
F-75014 Paris, France\\
$^3$Service de Physique Th\'eorique, C.E. de Saclay, 
F-91191 Gif-sur-Yvette, France}

\date{MNRAS 310, 428}




\maketitle

\begin{abstract} 
The errors on statistics measured in finite galaxy catalogs are
exhaustively investigated. The theory of errors on factorial moments
by Szapudi \& Colombi  (1996) is applied to cumulants via a series
expansion method. All results are subsequently extended to the weakly
non-linear regime.  Together with previous investigations this yields
an analytic theory of the errors for moments and connected moments of
counts in cells from highly nonlinear  to weakly nonlinear scales.
For nonlinear functions of unbiased estimators, such as the
cumulants, the
phenomenon of cosmic bias is identified and computed. 
Since it is subdued by the cosmic errors
in the range of applicability of the theory, 
correction for it is inconsequential.  
In addition, the method of Colombi, Szapudi \& Szalay (1998) concerning
sampling effects  is generalized,  adapting the theory for
inhomogeneous galaxy catalogs.  While previous work focused on the
variance only, the present article calculates the cross-correlations
between moments and connected moments as well for a statistically
complete description.  The final analytic formulae representing the
full theory are explicit but somewhat complicated.  Therefore as a
companion to this paper we supply a  FORTRAN program capable of
calculating  the described quantities numerically. An important
special case is the evaluation of the errors on
the two-point correlation function, for which this should be
more accurate than any method put forward previously.
This tool will be
immensely useful in the future both for assessing the precision of
measurements from existing catalogs, as well as aiding the design of
new galaxy surveys. To illustrate the applicability of the results and
to explore the numerical aspects of the theory qualitatively and
quantitatively, the errors and cross-correlations are predicted  under
a wide range of assumptions for the future Sloan Digital Sky
Survey. The principal results concerning the cumulants $\xiav$, 
$Q_3$ and, $Q_4$, is that the relative error is expected to be smaller than
3, 5, and 15 percent, respectively, in the scale range of $1\hmpc -
10\hmpc$; the cosmic bias will be negligible.

\end{abstract} 
\vskip 0.5cm

\section{Introduction}

According to theories of cosmological structure formation 
small initial fluctuations grew by gravitational amplification.
In the last decade, higher order statistics emerged 
as an important tool to test both the Gaussianity of initial conditions and
the gravitational amplification process. These tests are
a priori possible in the
perturbation theory (PT) regime where many predictions
have been obtained by now (see Juszkiewicz \& Bouchet 1995;
Bernardeau 1996b for recent short reviews), or in the nonlinear regime.
In both cases, they can potentially alleviate the ambiguity of 
the galaxy two-point correlation function
when light does not trace mass (biasing),
thereby shedding light on cosmology as well as 
the physics of galaxy formation (e.g. Fry \& Gazta\~naga 1993;
Gazta\~naga \& Frieman 1994; Szapudi 1998b).

A tight control of the errors is crucial
for the interpretation of higher order measurements
from galaxy catalogs. A sufficiently general and reliable
knowledge of the expected errors is all the more
timely as new galaxy surveys will come online in the near future.
Building on the  groundwork described in two previous papers,
Szapudi, \& Colombi (1996, hereafter SC), and Colombi,
Szapudi \& Szalay (1998, hereafter CSS),
the aim of this article is
to formulate a coherent analytic theory for the errors of
moments and connected moments of counts in cells in
all scale regimes for possibly inhomogeneous galaxy surveys.

There has been several explorations in the past concentrating
mainly on the errors of the two-point correlation function in
real and Fourier space
(e.g., Peebles 1980; Kaiser 1986; Landy \& Szalay 1993; Feldman,
Kaiser \& Peacock 1994; Hamilton 1993; Hamilton 1997a, 1997b;
Scoccimarro, Zaldarriaga \& Hui 1999)
or the $N$-point correlation functions \cite{ss98}. 
The analytic  calculation of the error on the void probability function
is described in Colombi, Bouchet \& Schaeffer (1995).
As moments of counts in cells have been the most successful descriptors
of higher order statistics so far, SC 
set out to formulate the general theory of variances
related to counts in cells in a finite galaxy catalog.
Explicit, analytic formulae were determined for estimating 
cosmic errors of the factorial moments. The main underlying
assumptions were the locally Poissonian approximation, and
the hierarchical ansatz for the higher order correlations.
The first consists of neglecting correlations among parts
of overlapping cells, while the latter is known
to be an excellent approximation in existing galaxy
catalog (e.g., Groth \& Peebles 1977; Fry \& Peebles 1978;
Sharp, Bonometto \& Lucchin 1984; Szapudi, Szalay \& Bosch\'an~1992;
Meiksin, Szapudi \& Szalay 1992; Bouchet et al.~1993;
Szapudi et al.~1995; Szapudi \& Szalay 1997)
and in $N$-body simulations in 
the highly non-linear regime (e.g., Efstathiou et al.~1988; Bouchet
et al.~1991; Bouchet \& Hernquist 1992; Fry, Melott \& Shandarin
1993; Bromley 1994; Lucchin et al.~1994; Colombi, Bouchet \&
Schaeffer 1994; Colombi, Bouchet \& Hernquist 1996; Munshi et al.~1999a;
Szapudi et al.~1999d).
CSS applied the previously developed theory
and investigated the effects of variable sampling
and thereby extended the results for inhomogeneous galaxy surveys.
An exhaustive description of the previous
calculations would be superfluous here since
all details can be found in SC.
Some of the main concepts and the general framework, however, is
summarized next. 

Careful examination of the generating
functions and their expansions yields a unique classification
of the errors according to their origin and an approximate
separation between them.
Part of the uncertainty on counts in cells
is due to the finite number of sampling cells, $C$. It is termed measurement
error and it is proportional to $1/\sqrt{C}$;
therefore it can be rendered arbitrarily
small. The algorithm of Szapudi (1998a) achieves
the limit of $C \rightarrow\infty$ in practice, i.e.~the measurement
errors are absent.

The rest of the variance, termed cosmic error, is
inherent to the galaxy catalog and cannot be substantially
improved upon except for extending the survey itself. It
splits further into a trichotomy of
finite volume effects,
arising from the fluctuations on scales larger than the survey,
edge effects, from the uneven weights given to galaxies in  
relation to survey geometry, and discreteness effects, due to the finite number
of galaxies tracing the underlying continuous random field. 
To leading order in $v/V$, these
three effects are approximately disjoint and the corresponding
relative errors are proportional to $[\xi({\hat L})]^{1/2}$, $(\xiav v/V)^{1/2}$,
and $[v/(V\bn^k)]^{1/2}$, respectively; $\xi({\hat L})$ is the integral
of the correlation function (with some restrictions) over the whole
survey area, $\xiav$ is the average correlation function in a cell,
$\bn$ is the average count in a cell, $k$ is the order of the statistic, and
$v$ and $V$ are the volumes of the cell
and the survey, respectively. Only the discreteness error depends
on the number of particles, and it disappears in the continuum limit.
The separation of these effects is only approximate, and depends
on the leading order nature of the calculation. Next to leading
order contributions are presented elsewhere (Colombi et al.~1999a).

There are further refinements and qualifications to the above summarized
theory. Edge effects, usually dominant on large scales, 
can be corrected for to some extent \cite{ls93,ss98}.
Such a correction  is always equivalent to a virtual extension of the survey,
thus it is controversial as often pointed out by ``fractalists''.
A fraction of discreteness effects depends on the geometry of the
survey thus can be termed as edge-discreteness effect
\cite{ss98}. Finally, finite volume effects overlap slightly with 
edge effects, even though the appropriate splitting of the 
corresponding integral yields an approximate separation.

The present work generalizes the previous calculations for
many useful statistics, such as the connected moments or cumulants
of the probability distribution of counts in cells, and extends
the validity of the theory into the weakly non-linear regime
by dropping the hierarchical assumption. Moreover, cross
correlation matrices for moments and connected moments are computed
as well for statistical completeness. To facilitate the 
practical application of this somewhat complicated 
but fully explicit and analytic theory, we supply 
FORTRAN programs to evaluate all the (co)variances
of moments and cumulants. This should diminish the
efforts needed to assess the accuracy of counts in 
cells measurements in present and future galaxy
catalogs, such as the Sloan Digital Sky Survey (SDSS) and
the two degree Field Survey, as well as in simulations.
In addition, design of future galaxy catalogs should be
optimized in light of the expected errors for 
different alternatives.  To demonstrate the practicality of our approach, 
the theory is illustrated throughout this paper 
by calculating the cosmic errors, cross-correlations, and biases for
all relevant statistics related to count-in-cells in the future SDSS.
It is worth to emphasize that our technique can be used to obtain
the errors on the two-point correlation function with more
accurate results than any previous method.

The next Section  describes the general theory of non-linear
error propagation including the resulting bias and the
calculation of (co)variances, with extension
of the analysis to the weakly non-linear regime. 
Sect. 3 presents practical results for the SDSS survey:
the expected errors, biases and cross-correlation of
factorial moments and cumulants up to fourth order are given 
for a wide variety of clustering models. Finally, Sect. 4 summarizes
and discusses the results. In addition, Appendix A 
illustrates the theory with explicit formulae 
too cumbersome to be included in the main text.
Appendix B compares in detail our predictions
for the cosmic bias on cumulants 
with the recent results of Hui \& Gazta\~naga 
(1998, hereafter HG).
%
%
\section{Theory}
%
%

In this section we present the theory of cosmic errors on the
quantities of interest, cumulants (or connected moments) $\xiav$ and 
$Q_N$ of the probability distribution function of the cosmic density.
The central issue addressed here is the propagation of errors
from the factorials moments $F_k$ to the cumulants, the latter being
{\it nonlinear} combinations of the former.
For this sake in
Sect.~2.1 we present the theory of error propagation in a general setting 
for functions of correlated random variables. 
Sect.~2.2 applies this formalism
to factorial moments and cumulants, taking advantage
of the theory of cosmic errors 
on factorial moments by SC.  Finally, Sect.~2.3  discusses 
the specific models of clustering employed for numerical
demonstration of the theory, including
generalization of the original framework for PT.

\subsection{General Error Propagation and Bias}

Let us assume that $f(x)$ is constructed from unbiased measurements of
a set of random variables $\{x_k\}$ with known errors and 
cross-correlations\footnote{As
long as errors on $x_k$ are small they can follow any
joint distribution. In particular they do not have to be Gaussian
distributed.}.
For measurements of a statistical quantity $x_k$, a different notation
(such as  $\tilde x_k$...) could be introduced for added precision.
However, such notation is dispensed of since
it would only clutter the formulae without adding
anything of importance.
If the measurements $\{x_k\}$ are sufficiently close to their ensemble
average, $\{\avg{x_k}\}$, it is meaningful to expand $f$ around
the mean value
\begin{equation}
  f(x)  = f(\avg{x}) + \frac{\partial f}{\partial x_k} \delta x_k +
\frac{1}{2}\frac{\partial^2 f}{\partial x_k\partial x_l} 
\delta x_k \delta x_l + \ldots {\cal O} (\delta x^3),
\end{equation}
where 
$\delta x_k = x_k - \avg{x_k}$, and the Einstein convention
was used.  It is fruitful 
to evaluate the variance and bias of $f$, and the cross-correlation of two
such functions $f,g$, up to second order precision.
The resulting theory will be reasonably
accurate as long as the variances and correlations of the
underlying statistics are
sufficiently small, 
i.e.~$\avg{\delta x_k \delta x_l}/\avg{x_k}\avg{x_l} \ll 1$.
Taking the ensemble average 
of the above equation  yields the average of $f$
in a finite survey
\begin{equation}
  \avg{f} = f(\avg{x}) + \frac{1}{2}\frac{\partial^2 f}
  {\partial x_k\partial x_l} \avg{\delta x_k \delta x_l} +
  \ldots {\cal O} (\delta x^3).
  \label{eq:errorexpansion}
\end{equation}
According to this equation $f(x)$ is a biased estimator
of $f(\avg{x})$  (see also HG). 
More precisely, if $x$ is an unbiased estimator, 
the (relative) bias on
$f(x)$ can be defined as
\begin{equation}
  b_f = \frac{\avg{f(x)} - f(\avg{x})}{f(\avg{x})}.
  \label{eq:biasdef}
\end{equation}
To second order, an unbiased estimator can be
constructed from the formula.
The bias is the result of the non-linear construction of
$f$ from unbiased measurements $x$. As the survey
becomes larger the errors decrease,
$\avg{\delta x_k \delta x_l} \rightarrow 0$, and, 
in agreement with intuition,  $f$ becomes less and less biased. 

Similarly the covariance
of two functions $f$ and $g$ can be evaluated,
\begin{equation}
  {\rm Cov}(f,g) = \avg{\delta f\delta g} = 
  \frac{\partial f}{\partial x_k}\frac{\partial g}{\partial x_l}
  \avg{\delta x_k \delta x_l}
  + {\cal O} (\delta x^3), \label{eq:qerror}
\end{equation}
where $\delta X = X -\avg X$.
The variance of a function $f$ is simply $(\Delta f)^2 \equiv {\rm Cov}(f,f)$,
and the relative error 
\begin{equation}
	\sigma_f = \sqrt{{\rm Cov}(f,f)}/\avg{f}=\frac{\Delta f}{\avg{f}}.
  \label{eq:sigsig}
\end{equation}
This is the general form of the widely quoted ``error propagation''
formula with correlated errors.

For a set of (possibly biased) statistics 
$f = \{f_k\}_{k=1,K}$, 
the covariance matrix is defined as $C_{ij} = {\rm Cov}(f_k,f_l)$,
which is in turn crucial for maximum likelihood analyses.
For reference, the
appropriate likelihood function in the Gaussian limit is 
(the logarithm of)
\begin{equation}
  \Upsilon(f) = \frac{1}{\sqrt{(2\pi)^K {\rm Det}(C)}} 
  \exp\left[ -\frac{1}{2} \delta f_k C^{-1}_{kl} \delta f_l \right],
\end{equation}
where ${\rm Det}(C)$ and $C^{-1}$ are the determinant and inverse
of the covariance matrix.

The range of applicability of the previous equations merits some
comments. The most obvious condition is that the 
relative (co)variance (\ref{eq:sigsig}), is $\sigma_f \ll 1$,
otherwise the Taylor expansion diverges.
From equations (\ref{eq:errorexpansion}), (\ref{eq:biasdef}) and
(\ref{eq:qerror}), the bias is of order $b_f  = {\cal O} (\sigma_f^2)$.
Clearly, there is a meaningful regime
\begin{equation}
b_f \ll \sigma_f \ll 1,
\label{eq:validity}
\end{equation}
where the theory is certainly valid.
In practice $b_f \simeq \sigma_f \ll 1$ can happen, 
contradicting, however,  the condition that $b_f \simeq \sigma_f^2$.
This is a sign of cancellations in the
coefficients, and in that case higher order expansions would
be necessary to obtain the leading order results. 

\subsection{Cosmic Errors and Cross-Correlations on Cumulants and
Factorial Moments}

For the present applications of the above formulae, 
the average count $\bn$, 
the variance $\xiav$, 
and the cumulants $Q_3$ and $Q_4$
are substituted for $\{f_k\}$. As
shown below, each of these can be expressed in terms
of the factorial moments $F_k$ (identified with $x_k$).
For further reference we first recall basic definitions,
then we formulate the theory of errors of SC
for factorial moments.

The variance of count in cells is the average of the correlation function in a cell
\begin{equation}
   \xiav \equiv \displaystyle \int {\rmd^3 r_1\over v} {\rmd^3 r_2\over v} 
           \xi(r_1,r_2).
\end{equation}
The cumulants of higher order are geometrical averages of the 
$N$-point correlations functions
\begin{equation}
   Q_{N} \equiv \frac{1}{N^{N-2}\xiav^{N-1}} \int
  \xi_{N}(r_1,\ldots,r_N) {\rmd^3r_1\over v}\ldots {\rmd^3r_N\over
  v},
\end{equation}
and by definition $Q_1\equiv Q_2 \equiv 1$.
Another widespread notation 
 exists in the literature for $Q_N$
\begin{equation} 
   S_N\equiv N^{N-2}\,Q_N,
\end{equation}
where the $N^{N-2}$ factor corresponds to
the number of trees that connect $N$ points.

The connected moments are non-linear functions of the factorial
moments (see Szapudi \& Szalay 1993), e.g., 
\begin{eqnarray}
  \bn &=& F_1 \\
  \xiav &=& \displaystyle \frac{F_2}{F_1^2} - 1  \\
  Q_3 &=& \displaystyle \frac{F_1 
\left(F_3-3 F_1 F_2+2 F_1^3\right)}{3(F_2-F_1^2)^2}\\
  Q_4 &=& \displaystyle \frac{F_1^2
\left(F_4-4 F_3 F_1-3F_2^2+12F_2 F_1^2-6 F_1^4\right)}{16(F_2-F_1^2)^3}, 
\end{eqnarray}
where
\begin{equation}
   F_k \equiv \langle (N)_k \rangle \equiv \langle N(N-1)\ldots(N-k+1)
   \rangle.
\end{equation}
Factorial moments are estimated in an unbiased fashion,
bias affecting the cumulants is due to non-linear construction.
Both the errors and biases of cumulants can be deduced
from the errors and cross-correlations 
of the factorial moments, 
${\rm Cov}(F_k,F_l)=\avg{\delta F_k \delta F_l}$, 
through the series expansions
(\ref{eq:errorexpansion}) and (\ref{eq:qerror})
if the variances are sufficiently small.

The diagonal term, ${\rm Cov}(F_k,F_k)$, was evaluated by SC under the
hierarchical and local Poisson behavior assumptions. 
For the present generalizations i) the hierarchical 
assumption has to be discarded (see next subsection), ii)  the $k \ne l$
cross terms need to be evaluated as well.
The cross-correlations of the factorial moments are obtained through
a completely analogous if cumbersome calculation as described in
SC. The basic steps are outlined next.

To evaluate the cross-correlations the full error generating
function of the factorial moments,  which contains the measurement 
errors and the cosmic errors, should be expanded (SC),
\begin{equation}
  {\rm Cov}(F_k,F_l)=\left. \left[ \frac{\partial}{\partial x_{}} \right]^k
\left[ \frac{\partial}{\partial y} \right]^l E^{C,V}(x+1,y+1)
\right|_{x=y=0},
\end{equation}
\begin{equation}
  E^{C,V}(x,y) = \left(1 -\frac{1}{C} \right) E^{\infty,V}(x,y)
  +E^{C,\infty}(x,y). 
  \label{errortot}
\end{equation}

For completeness,  the measurement errors are generated by
\cite{sc96}, 
\begin{equation}
   E^{C,\infty}(x,y) = \frac{\Gen(xy)-\Gen(x)\Gen(y)}{C},
\end{equation}
where $\Gen(x)$ is the generating function of the 
distribution of counts in cells; 
$F_k=(\rmd/\rmd x)^k \Gen(x+1)|_{x=0}$.
The measurement errors can always be eliminated with
large or infinite number of sampling cells employed 
in state of the art measurement algorithms \cite{sz97,sqsl98}.
Therefore the limit $C \rightarrow \infty$ is taken, i.e. the
number of sampling cells tends to infinity, and measurement
errors shall not be mentioned further.

The surviving part of the generating function is $E^{\infty,V}(x,y)
= \avg{\Gen(x)\Gen(y)} - \avg{\Gen(x)}\avg{\Gen(y)}$ with
\begin{equation}
  \avg{\Gen(x)\Gen(y)} \equiv
  {1\over {\tV}^2} \int_{\tV} \rmd^Dr_1 \rmd^Dr_2 \Gen(x,y) = 
  \int_{\rm o}+\int_{\rm no},
  \label{eq:error}
\end{equation}
where $D$ is the dimension of the survey, ${\hat V}$ is the
volume covered by cells included in the catalog and $\Gen(x,y)$ is
the generating function of bicounts for cells separated by
a distance $|r_1-r_2|$. Throughout the
paper three-dimensional geometry is assumed. The above
equation yields both cosmic errors and cross-correlations.
The calculation is facilitated by separating the
double integral
according to whether cells corresponding to coordinates
overlap (o) or not (no). Details can be found in SC
where $k=l$ terms were evaluated. 

The contribution to the cosmic errors from disjoint cells 
corresponds to the finite volume errors, obtained 
from Taylor expanding the bivariate generating function
of counts in cells, as shown below. 

The contribution from overlapping cells corresponds to the
edge and discreteness effects. Its evaluation
is  somewhat tedious, involving a numerical integration 
after the expansion of the generating function. Nevertheless 
there are no further complications compared to the diagonal case
of SC. The locally Poissonian assumption allows a major
simplification of the calculation: only the
monovariate generating function is integrated instead
of the significantly more complicated trivariate function.

\subsection{Generating functions and models}

The original calculations of SC were based on 
a successful model for the highly non-linear regime,
the hierarchical tree assumption. This assumption has never
been fully demonstrated although some hints for it has been
given recently (e.g., Scoccimarro \& Frieman 1998).
Since the coherent infall on large scales introduces an angle
dependence in the perturbation theory kernels 
(e.g., Goroff et al.~1986), 
this approximation breaks down in the weakly
non-linear regime. This necessitated a generalization
of the previously used assumptions for this article. 
The resulting new generating functions
accommodate most models currently used, such as the Ansatz by
Szapudi \& Szalay (1993), denoted by SS and
the one by Bernardeau \& Schaeffer (1992), 
denoted by BeS, perturbation theory (PT), 
and extended perturbation theory (Colombi et al.~1997),
hereafter EPT.

The other simplifying assumption of SC, the
local Poissonian Ansatz, is kept for the present calculations.
To eliminate it would require major modification in the numerical method,
due to the trivariate generating function. 
Fortunately all indications point to the extreme accuracy of 
this assumption for error calculations, although for cross
correlations it becomes increasingly questionable as
the difference of orders, $|k-l|$, increases \cite{hubble1}.

Since the models and the method of calculation are described
by SC in sufficient detail, only the new features
arising from the present general setting 
are pointed out next. 

As described in \S~2.2, the calculation of errors requires
the knowledge of the monovariate and the bivariate generating
functions for the counts.

The monovariate generating function remains
formally unchanged compared to SC, since the original form 
(White 1979; Balian \& Schaeffer 1989; Szapudi \& Szalay 1993) is
completely general, 
\begin{equation}
   \Gen(x) = \exp\left\{ \sum_{N=1}^\infty (x-1)^N\Gamma_N Q_N \right\},
          \label{eq:genpn}
\end{equation}
with 
\begin{equation}
    \Gamma_N = {N^{N-2} \over N!} \bn^N\ \xiav^{N-1}.
\end{equation}
However, various assumptions about $\xiav$ and $Q_N$ 
are different for each model. These can be obtained either 
from measurements or phenomenology in the case of the SS and 
BeS models, or from the form of the primordial power spectrum for PT.
PT has specific rules to relate $Q_N$ to the local
derivatives of the power spectrum (Juszkiewicz, Bouchet \& Colombi
1993; Bernardeau 1994a,b), e.g., 
\begin{eqnarray}
  Q_3 & = & \displaystyle \frac{34}{21} - \frac{n+3}{3}, \label{eq:q3} \\
  Q_4 & = & \displaystyle \frac{7589}{2646} - \frac{31(n+3)}{24} +
        \frac{7(n+3)^2}{48}\label{eq:q4},
\end{eqnarray}
with $n = -3-\rmd\log\xiav/\rmd\log\ell$. From here on
higher order derivatives 
$\gamma_j = \rmd^j \xiav/(\rmd\log\ell)^j$ are
neglected in the calculation
of $Q_N$, $N\geq 4$ (Bernardeau 1994b). 
This is an accurate
approximation and simplifies the calculations 
(e.g., Colombi et al.~1999b).

The general form of the bivariate generating function is
(Schaeffer 1985; Bernardeau \& Schaeffer 1992; Szapudi \& Szalay 1993)
\begin{equation}
     \Gen(x,y) =\Gen(x)\Gen(y)\exp\left[ R(x,y) \right], 
\end{equation}
where $R(x,y)$ contains the cumulants 
connecting two cells,
\begin{equation}
R(x,y)= \displaystyle \xi\sum_{M=1,N=1}^\infty (x\!-\!1)^M(y\!-\!1)^N 
              Q_{NM}\Gamma_M\Gamma_N NM. 
            \label{eq:pxy}
\end{equation}
The coefficients $Q_{NM}$, the
cumulant correlators, are defined similarly to the
$Q_N$'s,
\begin{eqnarray}
   Q_{NM} &=& \frac{1}{N^{N-1}M^{M-1}}\frac{1}{\xiav^{N+M-2}}\nonumber\\
   &&\times\int_{v_1,v_2} \xi_{N+M} {\rmd^3r_1\over v}\ldots
   {\rmd^3r_{N+M}\over v}.
\end{eqnarray}
This is an integral of the $N+M$-point correlation function over two 
separate cells. The normalization corresponds to the number
of possible trees in each cell multiplied with possible non-loop
connections between the cells multiplied with the appropriate
power of the average correlation function. Thus the $Q_{NM}$'s 
become unity when the underlying
tree graphs of the higher order correlation functions are
all given unit weights \cite{schaeff85,ss97}. Note the alternative
notation $C_{NM}  = Q_{NM} N^{N-1} M^{M-1}$ \cite{bern96}.

When the cell separation is much larger than the cell radius it is
natural to expand the generating function in terms 
of $\xi/\xiav$ \cite{bs92,ss93,sc96,ss97}.
As a consequence $\exp(R[x,y]) \simeq 1+R(x,y)$, thus
\begin{equation}
\Gen(x,y)
\simeq\Gen(x)\Gen(y)\left[ 1+R(x,y) \right] + {\cal O}(\xi^2/\xiav^2).
\label{eq:rexp}
\end{equation}
The above was found to be extremely accurate
in practice, even for touching cells.

Phenomenological theories of the bivariate counts attempt to
relate the cumulant correlators, $Q_{NM}$,  to the cumulants, $Q_N$. 
The leading assumptions, used for the numerical explorations
of Sect. 3, are reviewed next.

The SS approximation is purely phenomenological.
It assumes that  
\begin{equation}
  Q_{NM}^{\rm SS} = Q_{N+M}.
\end{equation} 
For example
\begin{eqnarray}
   Q_{12}^{\rm SS} & = & Q_3,  \nonumber \\
   Q_{13}^{\rm SS} & = & Q_{22}^{\rm SS} \, = \, Q_4.
\end{eqnarray}

The BeS model postulates a factorization
property for the joint counts in cells, $P_{NM}$. 
From  equation (\ref{eq:rexp}),
\begin{equation}
P_{NM}\simeq P_{N}\,P_{M}\left(1+b_{NM}\,\xi\right).
\label{eq:Pfact}
\end{equation}
In addition the BeS model imposes\footnote{This is also suggested by recent
numerical results obtained by Munshi, Coles \& Melott (1999b) in 2D dynamics.} 
that $b_{NM} = b_{N}b_{M}$, implying
\begin{equation}
Q^{\rm BeS}_{NM}=Q^{\rm BeS}_{N1}\,Q^{\rm BeS}_{M1}.
\label{eq:Qfact}
\end{equation}
This is true in a minimal tree construction
providing specific relationships between $Q_N$'s and
$Q_{N1}$'s (see Bernardeau \& Schaeffer 1992 and Bernardeau \&
Schaeffer 1999 for a more detailed discussion of this model). 
For instance,
\begin{eqnarray}
   Q_{12}^{\rm BeS} & = & Q_3,\nonumber   \\
   Q_{13}^{\rm BeS} & = & \frac{4}{3}Q_4-\frac{1}{3}Q_3^2, \nonumber \\
   Q_{22}^{\rm BeS} & = & Q_3^2.\label{eq:QijBeS}
\end{eqnarray} 
Interestingly, the SS and BeS models are identical when $Q_N=1$ for
all $N$. Since in practice, the $Q_N$'s depart from unity only
weakly, the difference between the two models is usually insignificant,
despite the formal dissimilarity between them.

When calculations are done in PT framework the properties 
(\ref{eq:Pfact}) and (\ref{eq:Qfact}) are 
also naturally obtained \cite{bern96},
\begin{equation}
Q^{\rm PT}_{NM}=Q^{\rm PT}_{N1}\,Q^{\rm PT}_{M1}.\label{eq:QfactPT}
\end{equation}
The evaluation of the lowest non-trivial orders yields
(Fry 1984; Bernardeau 1996),
\begin{eqnarray}
   Q_{12}^{\rm PT} & = & \displaystyle \frac{34}{21} - \frac{n+3}{6}, \nonumber \\
   Q_{13}^{\rm PT} & = & \displaystyle \frac{11710}{3969} - \frac{61(n+3)}{63}
             +\frac{2(n+3)^2}{27}, \nonumber \\
   Q_{22}^{\rm PT} & = & \displaystyle \frac{1156}{441} -
   \frac{34(n+3)}{63} + \frac{(n+3)^2}{36},
\label{eq:q21}
\end{eqnarray}
where $\gamma_2= \rmd^2 \xiav/(\rmd\log\ell)^2$
term in the second equation above is neglected as previously.

Note that, in the weakly nonlinear regime where the $Q_N$'s
are given by equations (\ref{eq:q3}) and (\ref{eq:q4}), 
SS and BeS models give factors $Q_{NM}$
of same order as the correct result (\ref{eq:q21}). In fact, the
BeS model agrees exactly with PT for $n=-3$. 

PT as a model can be extended throughout
the non-linear regime as well. In the resulting theory,
EPT (Colombi et al.~1997), the form of the $Q_N$'s is still
taken from PT; e.g., equation (\ref{eq:q3})
can be extended into the non-linear regime. Then $n$,
formerly the slope of the power spectrum,
becomes a formal fitting parameter, denoted with $n_{\rm eff}$.
It was found empirically in simulations and galaxy 
data that {\em all} higher order
$Q_N$ can be described fairly accurately with a single $n_{\rm eff}$ 
parameter \cite{cbbh97,smn96,sqsl98}. This idea
can be generalized to the bivariate distribution in several
ways, as proposed by Szapudi \& Szalay (1997). The version used in this
work, denoted by E$^2$PT, consists of taking
the same $n_{\rm eff}$ for the $Q_{NM}$'s in equations (\ref{eq:q21}) 
as for the $Q_N$'s.

The new assumptions for the generating function are sufficiently
general to incorporate most conceivable models, notably
perturbation theory and its variants. Fortunately the
changes do not incur many complications for the error calculations
compared to that of SC.
The overlapping part of the integral  in equation (\ref{eq:error})
depends on the unchanged monovariate distribution.
This calculation, the most complicated and CPU consuming component
of the technique, was performed by SC. Here
only the appropriate 
values of $\xiav$ and 
the $Q_N$'s had to be substituted into the analytic results. 
The missing cross-correlations of the overlapping part
were computed in an exactly analogous fashion as
previously. This somewhat cumbersome task was carried out
by the {\tt Mathematica} computer algebra package.

The bivariate generating function induces
the non-overlapping part of the integral constituting the 
error generating function,
i.e. the finite volume effects. The computation 
consists of expanding  equation (\ref{eq:pxy}), 
a simple albeit tedious analytical computation
performed again with {\tt Mathematica}.

Up to the locally Poissonian assumption and the expansion
of the bivariate generating function 
to linear order in $\xi$ (an excellent approximation even for touching
cells), the results are completely general, and can be
used easily if new interesting models surface. 

Explicit analytic expressions for the cosmic errors and cross-correlations
are given in Appendix A for factorial moments, up to third order. 
\section{Application: SDSS-like surveys}

The results were applied to calculate the expected errors,
cross-correlations, and biases for SDSS-like galaxy
catalogs as defined in detail in CSS.
The SDSS is a magnitude limited galaxy survey where
the average number density of galaxies decreases with distance
from the observer.
To investigate a reasonable range of underlying clustering properties,  
the shape and normalization of the two-point correlation function, 
thus $\xiav$ and $\xi({\hat L})$ (see introduction and Appendix A)
were taken from the standard CDM model of CSS (hereafter CDM1) as well
as four CDM variants proposed by the Virgo Consortium
\cite{j98} (hereafter CDM2,3,4,5, as described in Table~1). 
\begin{table}
\caption{The standard CDM model used by CSS (CDM1) and the four 
CDM variants proposed by the Virgo Consortium (CDM2,3,4,5).
The notations are the same as in Jenkins et al.~(1998).}
\begin{tabular}{lccccc}
\hline
Model & $\Omega_0$ & $\Lambda$ & $h$   & $\Gamma$ & $\sigma_8$  \\ \hline
CDM1  & $1.0$        & $0.0$     & $0.5$ & $0.50$   & $1.00$ \\
CDM2  & $0.3$        & $0.0$     & $0.7$ & $0.21$   & $0.85$ \\
CDM3  & $0.3$        & $0.7$     & $0.7$ & $0.21$   & $0.90$ \\
CDM4  & $1.0$        & $0.0$     & $0.5$ & $0.21$   & $0.51$ \\
CDM5  & $1.0$        & $0.0$     & $0.5$ & $0.50$   & $0.51$ \\ \hline
\end{tabular}
\end{table}
CDM1 is used as default, except when otherwise indicated.
The SS and BeS models depend on the higher order cumulants $Q_N$
thus EPT could be used with
$n = -2.5$. This agrees approximately with the
measurements in the APM and EDSGC \cite{gaz94,sdes95,smn96,sg98}.
The same spectral index was used as default for E$^2$PT, as well
as the indices $n  = -1$ and $n = -9$ for reasonable 
alternatives of higher order clustering, especially in the highly
non-linear regime. The most successful model
of all for error calculations \cite{hubble1}, E$^2$PT was
used as a default unless otherwise noted.

For the sake of conciseness, the technical information
on figures is contained in the captions only and the
physical results are explained in the main text {\em with the least
possible overlap}. The more conventional procedure of 
duplicating information
in the main text would have rendered the paper unnecessarily long
and cumbersome due to the exceptionally large number of
figures and the multitude of line-types, panels, etc.  contained in them.

\subsection{Cosmic Errors and Bias}
\begin{figure*}
\centerline{\hbox{\psfig{figure=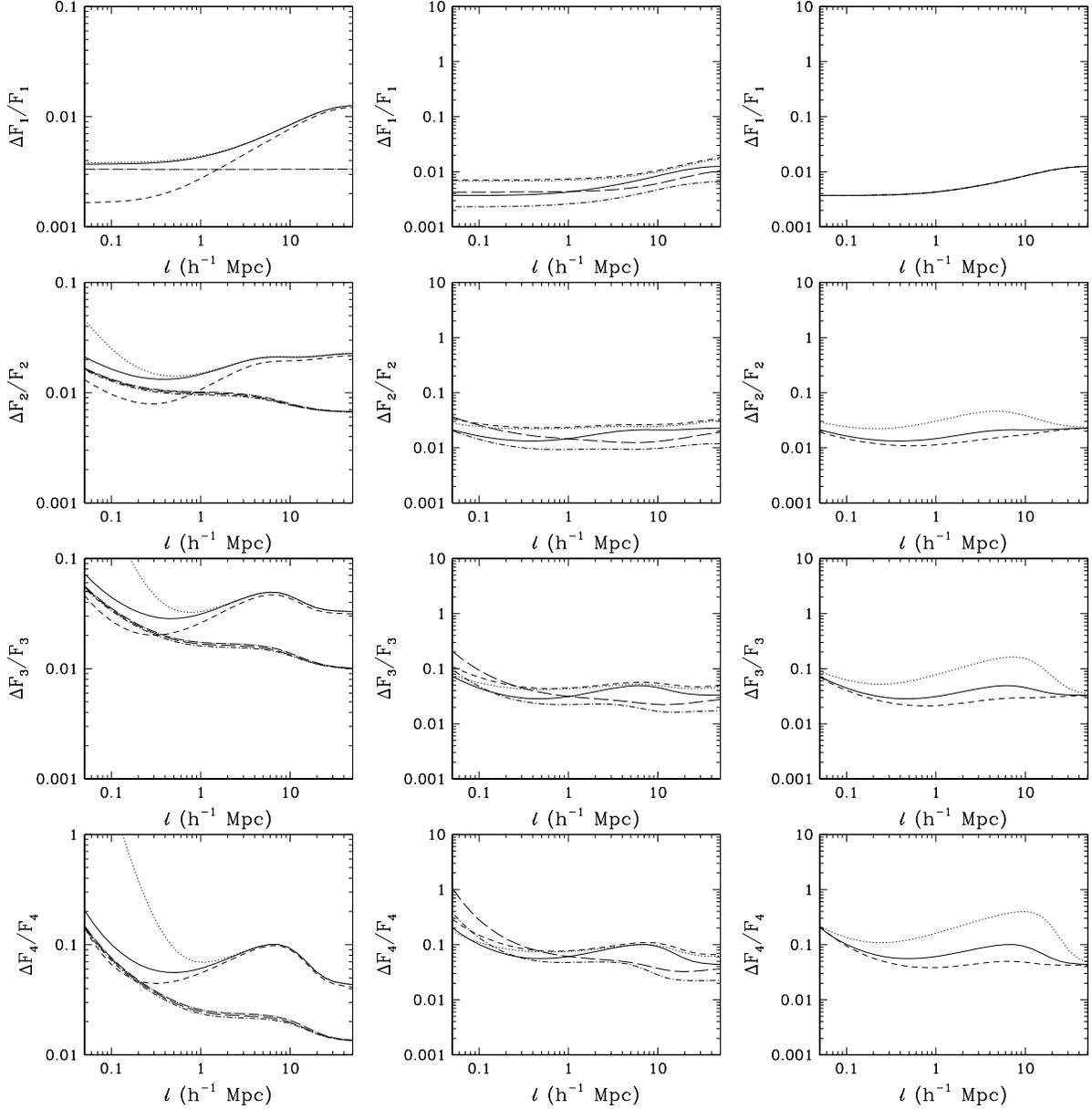,bbllx=28pt,bblly=237pt,bburx=570pt,bbury=800pt,width=16cm}}}
\caption[]{ Prediction of the cosmic error
on factorial moments, $\sigma_{F_k}=\Delta F_k/F_k$, for $k \le 4$. The first
column shows  the cosmic error from disjoint cells for
different models SS (long-dashes), BeS (dot-dashes), 
E$^2$PT (long dashes with dots), and also from overlapping
cells (dashes). The indistinguishable solid curves display the total error
for each model.
All the above assumes optimal radial sampling weight $\omega$
\cite{css98}, while the dotted line was computed with
uniform weight for comparison. The second and third column demonstrate
the robustness of the results with respect to 
variation of the two-point correlation function in
the different CDM models (respectively solid, dots, dashes, long
dashes and dot-dashes for CDM1,2,3,4,5), and the choice of the spectral
index for E$^2$PT (solid for $n=-2.5$, dots for $n=-9$ 
and dashes for $n=-1$), respectively.
In the first column $n = -2.5$ and CDM1 was used. Second
column has $n = -2.5$ with E$^2$PT,  the third column has CDM1 with
E$^2$PT. Note that for the first and third columns the errors
for $F_1$ are independent of higher order statistics, therefore
the different models superpose.}
\label{fig:figure1}
\end{figure*}

Figure 1 shows the expected errors on the factorial moments 
in SDSS-like surveys for various models and contributions.
The estimator for the factorial moments proposed by
CSS is assumed,
\begin{equation}
  \tF_k^C \equiv \frac{1}{C} \sum_{i=1}^C \frac{\left( N_i \right)_k 
  \omega_{\ell,k}(\r_i)}{[\phi_\ell(r_i)]^k},
  \label{eq:myw}
\end{equation}
where $C$ is the (very large) number of sampling cells 
thrown at positions $\r_i$, $\phi_\ell(r_i)$ is the selection function, and
the weight $\omega_{\ell,k}$ is determined to minimize
the variance of the estimator. As shown by CSS,
the weights can be optimized by numerically solving
an integro-differential equation, while the approximate
solution is $\omega \propto 1/\sigma^2$, with $\sigma$ representing
the full errors of the given statistic.
The above optimal weight is assumed for most curves. (See the
figure caption for details). In general, i) the different models
SS, BeS, and E$^2$PT yield almost same results, ii) the dependence
on the two point function causes a spread reaching a factor of $5$ on certain
scales almost independently of order, iii) 
different reasonable assumptions for the underlying $Q_N$'s 
generate significant spread which,
depending both on order and scale, can reach up to an order of magnitude. 
The assumptions for the $Q_N$'s, however, allowed a quite
generous variation taking into account the typical difference 
between weakly non-linear and highly non-linear regime in
CDM-type simulations. 
Uniform weighting scheme boosts the errors on small scales
considerably compared to the optimal weights introduced
by CSS except for $F_1$ where there is no significant
difference. In most of the relevant dynamic range, 
$1\hmpc \le \ell \le 50\hmpc$, edge effects
are dominating the errors. For any realistic survey, 
the geometry is expected to be more complex because
of the cut out holes caused by bright stars, cosmic rays, etc. This could
significantly boost the edge effects compared to the calculations
presented here. Discreteness effects are important
for very small scales $\ell \le 1\hmpc$ and uniform
weights only.  Optimal weights render discreteness 
and finite volume effects on a par in this regime.

\begin{figure*}
\centerline{\hbox{\psfig{figure=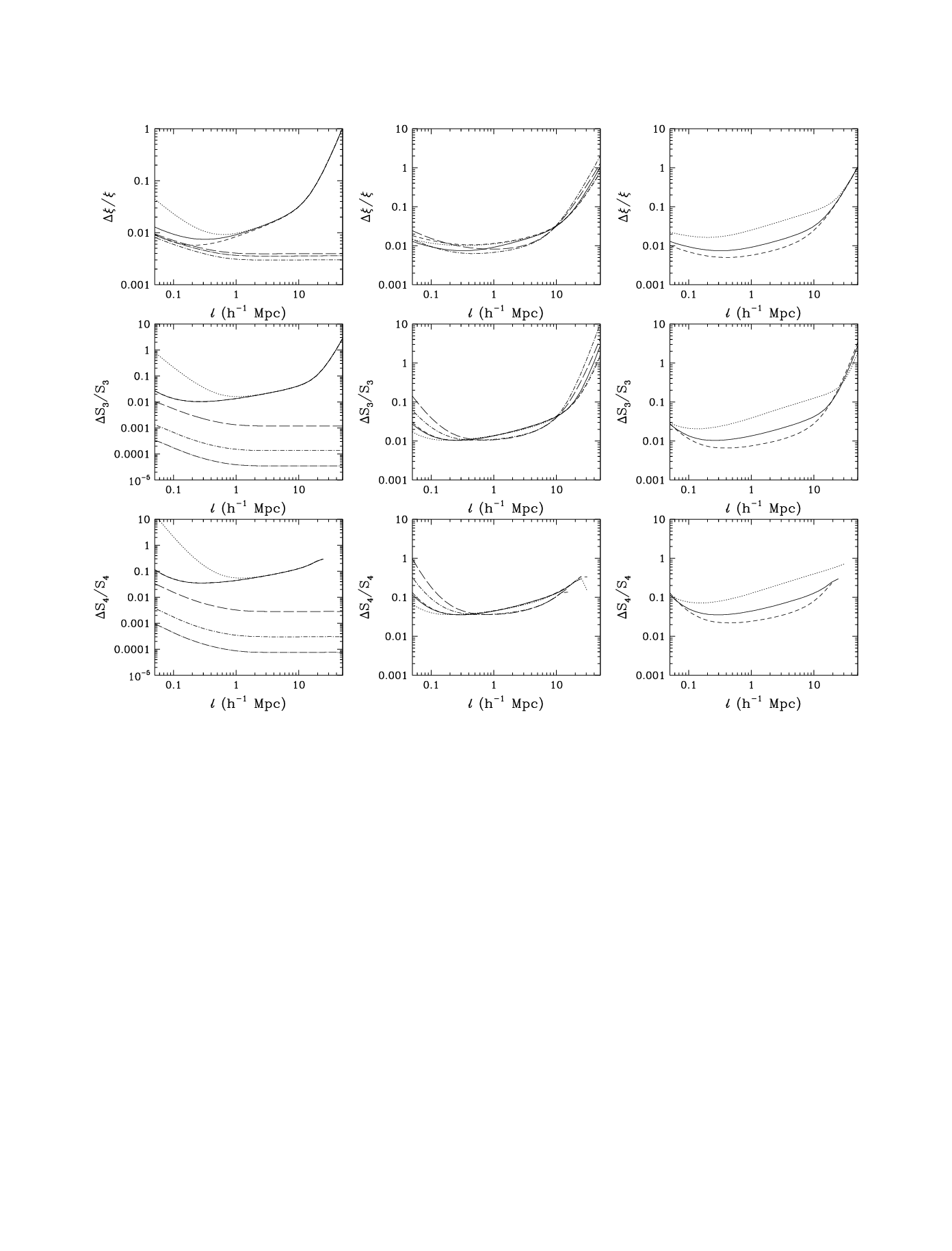,bbllx=23pt,bblly=376pt,bburx=570pt,bbury=800pt,width=16cm}}}
\caption[]{ Same as Fig.~1 for connected moments, i.e.
for $\xiav$ and $S_N = Q_N N^{N-2}$. The curves are only
plotted when the expansion in equation (\ref{eq:qerror})
yields positive results for the cosmic error.
}
\label{fig:figure2}
\end{figure*}

Figure 2 is analogous to Figure 1 for the connected moments.
In contrast with the factorial moments, i) finite volume error
is completely negligible compared to the other contributions, 
and for orders $N > 2$ it is strongly dependent on the 
models, SS $\gg$ BeS $\gg$ E$^2$PT, ii) the dependence on the
two point correlations is less pronounced,
 iii) the dependence on higher order clustering 
appears to be less sensitive to order.

\begin{figure}
\centerline{\hbox{\psfig{figure=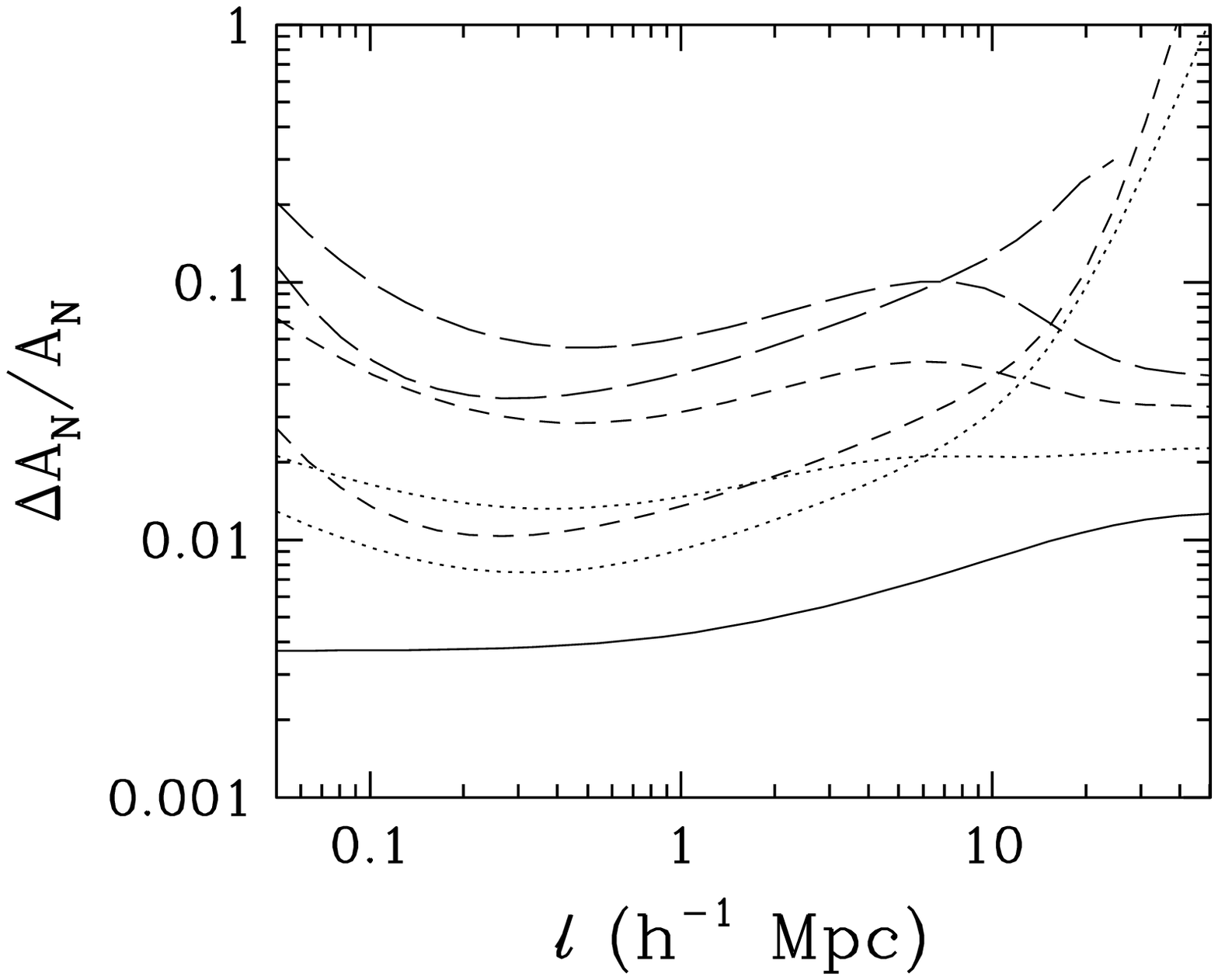,bbllx=48pt,bblly=112pt,bburx=521pt,bbury=488pt,width=8cm}}}
\caption[]{ Comparison of the cosmic errors for the factorial and connected
moments. CDM1 was assumed for the two-point correlation
function and E$^2$PT with $n = -2.5$ for higher order statistics. 
Solid, dotted, dash, and long dash lines correspond to orders $1$ through $4$,
respectively. Of each pair of curves with the same line-types
the one turning up on large scales relates to the cumulant.
The right stopping point of the long dash curve for
$S_4=16\,Q_4$ was determined similarly to Figure 2.
}

\label{fig:figure3}
\end{figure}

Figure 3 recapitulates the results of Figs. 1-2 by
comparing the errors on measurements of 
factorial moments with connected moments.
For small scales $\ell \la 7-10\hmpc$ the cumulants fare much better than
factorial moments; one reason is the suppression of
finite volume effects. Note especially the large difference between
$Q_3$ and $F_3$. Interestingly, 
$Q_3$  has small errors, within a factor of two $\Delta\xiav/\xiav$, and
there is a range in which $\Delta Q_3/Q_3 \la \Delta F_2/F_2$.
The edge effects for the cumulants are greatly boosted on large scales
compared to the factorial moments. However, this has to be 
interpreted cautiously since those scales are close to the limit of 
applicability of the theory according to equation (\ref{eq:validity}).

\begin{figure}
\centerline{\hbox{\psfig{figure=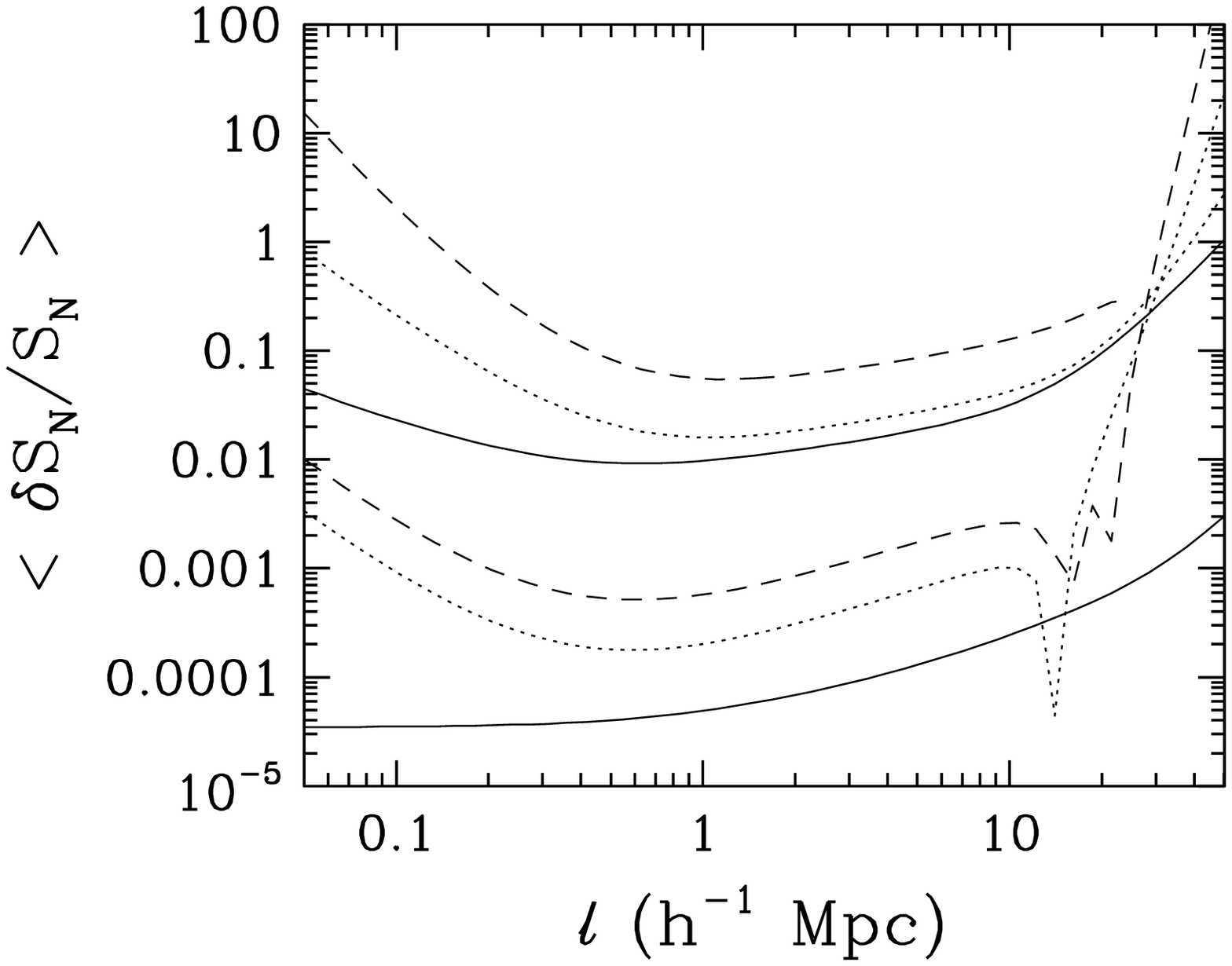,bbllx=48pt,bblly=112pt,bburx=521pt,bbury=488pt,width=8cm}}}
\caption[]{ The comparison of the cosmic bias and the
cosmic error for the cumulants. For all curves CDM1 and
E$^2$PT with $n = -2.5$ were used. Line-types correspond to
$\xiav$ (solid) , $S_3=3 Q_3$ (dotted) , and $S_4=16 Q_4$ (dashed), 
respectively. The three lower curves show the absolute value of the cosmic
bias, while the three upper ones correspond to the cosmic
error. The end point of the curve for $\Delta S_4/S_4$
was determined as previously (Fig.~2). For the cosmic 
bias $b_{Q_N}$ there is some irregularity
above $\sim 10\hmpc$. At this point the validity of our theory 
is probably exceeded, and the results become unstable.
In the regime where the theory is applicable the cosmic bias is 
always negative for the SDSS catalog. 
}
\label{fig:figure4}
\end{figure}

Figure 4 compares the magnitude of the cosmic bias to the cosmic error
for cumulants. As expected from theoretical prejudice, the cosmic
bias is by orders of magnitude smaller than the cosmic error 
in the regime where the perturbative
approach is applicable, i.e. $\ell \la 10\hmpc$ for the SDSS.
On larger scales the bias calculation apparently becomes unstable.
Thus Figure 4 re-confirms the correctness of 
equation (\ref{eq:validity}) as a guidance for the validity of the theory.

\subsection{Cosmic Cross-Correlations}
\begin{figure*}
\centerline{\hbox{\psfig{figure=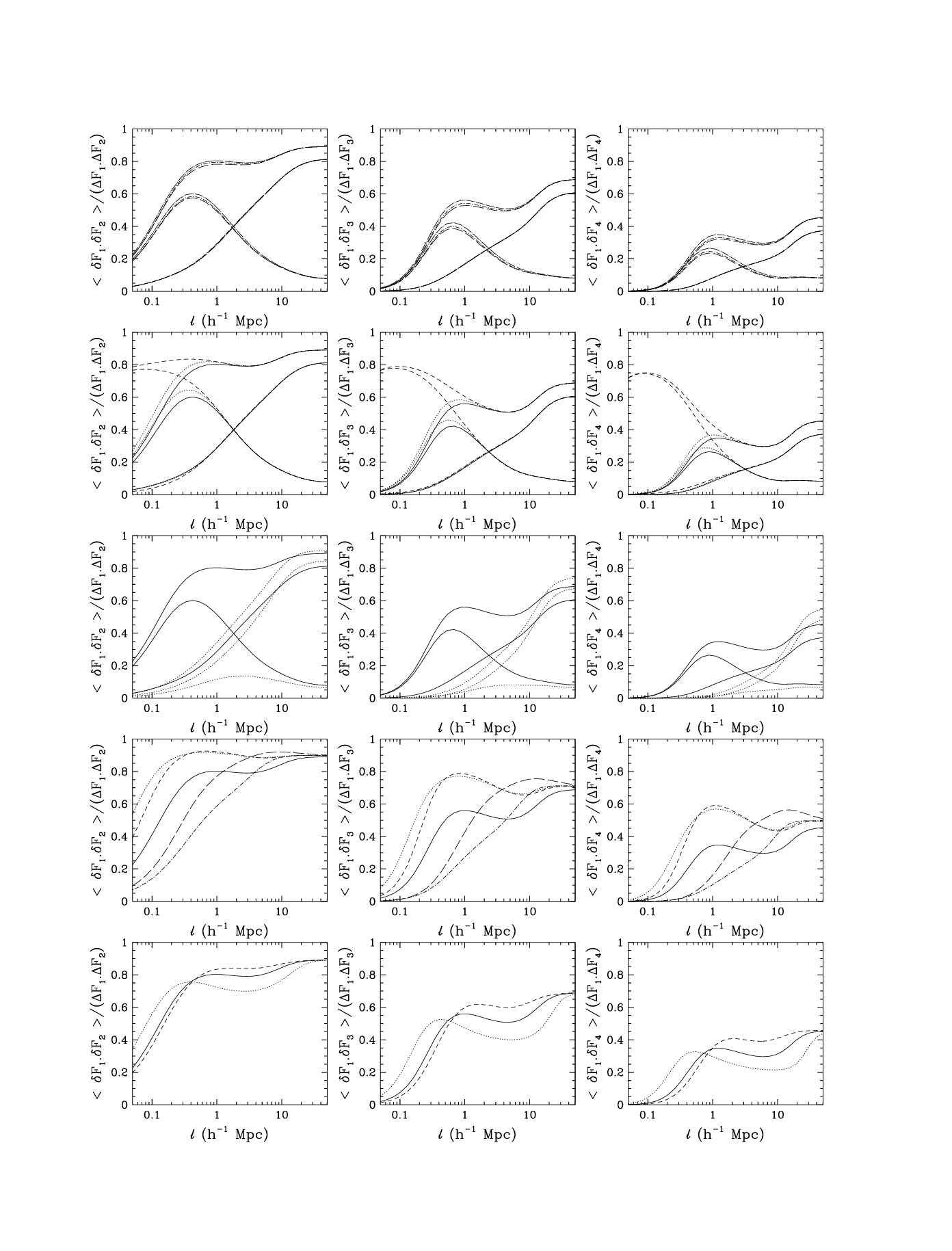,bbllx=22pt,bblly=73pt,bburx=550pt,bbury=800pt,width=13cm}}}
\caption[]{ Cosmic cross-correlation coefficients of the factorial moments.
The individual columns correspond to cross terms $\delta_{kl}$
for pairs of indices
$(k,l) = (1,2), (1,3), (1,4), (2,3), (2,4), (3,4)$, respectively.
Homogeneous weights were used for all panels, except for the second
row. Except for row five, $n = -2.5$ is assumed for higher order
statistics. Except for the first row, E$^2$PT is used. Finally,
except for the fourth row, CDM1 is the underlying cosmology.

The first row of panels compares various contributions
within the framework of  SS (long-dashes), BeS (dot-dashes), 
E$^2$PT (long dashes with dots). The difference between the 
three models is negligible.  
The resulting three groups
of curves in increasing order at $\ell = 8\hmpc$ 
correspond to the finite volume,
overlapping (i.e. discreteness+edge effects), and the total contributions,
respectively. Note that the full cosmic error was used in the denominator for
each curve to preserve additivity. This explains the residual
dependence of the overlapping contributions on the model.

The second row is analogous to the first one but examines the dependence 
on the optimal weights. The uniform weights (solid) are compared
to the optimal weights for orders $k$ (dots) and $l$ (dashes), where
$k < l$.

The third row illustrates dilution effects. The full sampling is shown
by solid lines while the effects of 10 times  dilution are displayed
by dots. The curves in increasing order at $\ell = 8\hmpc$ again
correspond to the finite volume,
overlapping (i.e. discreteness+edge effects), and the total contributions,
respectively.

The fourth row displays how total contributions are affected
by the choice of the two-point correlation functions in
different variants CDM1 through CDM5 (in the same order, 
solid, dots, dashes, long dashes, dot-dashes), respectively.

The fifth row shows the changes on the cross-correlations
due to varying the higher order statistics via
changing the spectral index in the framework of E$^2$PT, $n = -1$ (dashes) 
$n = -2.5$ (solid), and $n = -9$ (dots).

}
\label{fig:figure5}
\end{figure*}
\begin{figure*}
\centerline{\hbox{\psfig{figure=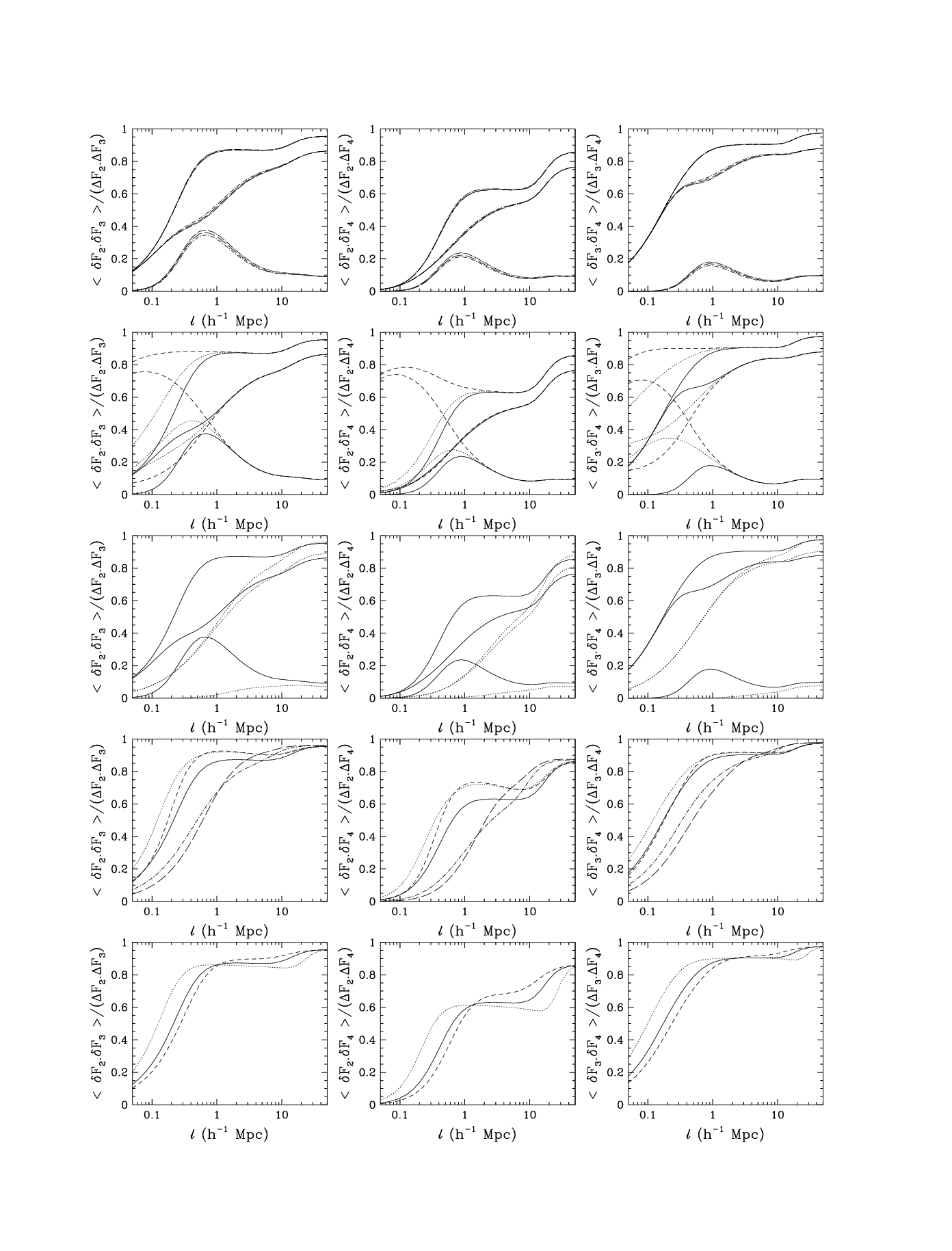,bbllx=22pt,bblly=73pt,bburx=550pt,bbury=800pt,width=13cm}}}
\nonumber
{\small {\bf \noindent Figure 5:} Continued.}
\end{figure*}

Figure~5 displays the cross-correlation coefficients 
\begin{equation}
  \delta_{kl}  = \frac{\avg{\delta F_k \delta F_l}}{\Delta F_k \Delta F_l}
  \label{eq:coeff}
\end{equation}
for factorial moments under various circumstances. In this equation
the denominator {\em always} contains the full cosmic error even
when only certain contributions are examined for the cross-correlations;
this ensures additivity.  For most
calculations homogeneous weights were used. The correlations
increase from small scales $\ell \la 1\hmpc$ to an approximate plateau.
The finite volume contribution exhibits a unimodal behavior
with a peak on small scales, while edge effects rise on large
scales. The shape of the finite volume part is
mainly due to the division by the full cosmic error in the
previous equation:
on small scales discreteness, on large scales edge effects
cause suppression. The same argument applies to
the drop of the full coefficient on small scales: discreteness (therefore
dilution) boosts the cosmic errors, thus reduces $\delta_{kl}$.
Note also that the relative contribution of the finite
volume effect is decreasing with order as already found for
the cosmic error. 

In addition to the previous comments, the following 
observations can be made from Fig.~5:
i) similarly to the cosmic errors, the different models SS, BeS, 
and E$^2$PT yield almost exactly the same cross-correlations, ii) 
optimal weighting naturally increases the cross-correlation, especially
on small scales and when the weights are selected to be optimal
for the higher order of the two statistics, iii) the effects of the choice of
the two-point correlation function are considerable, while 
the results are robust against variations of higher order statistics.

\begin{figure*}
\centerline{\hbox{\psfig{figure=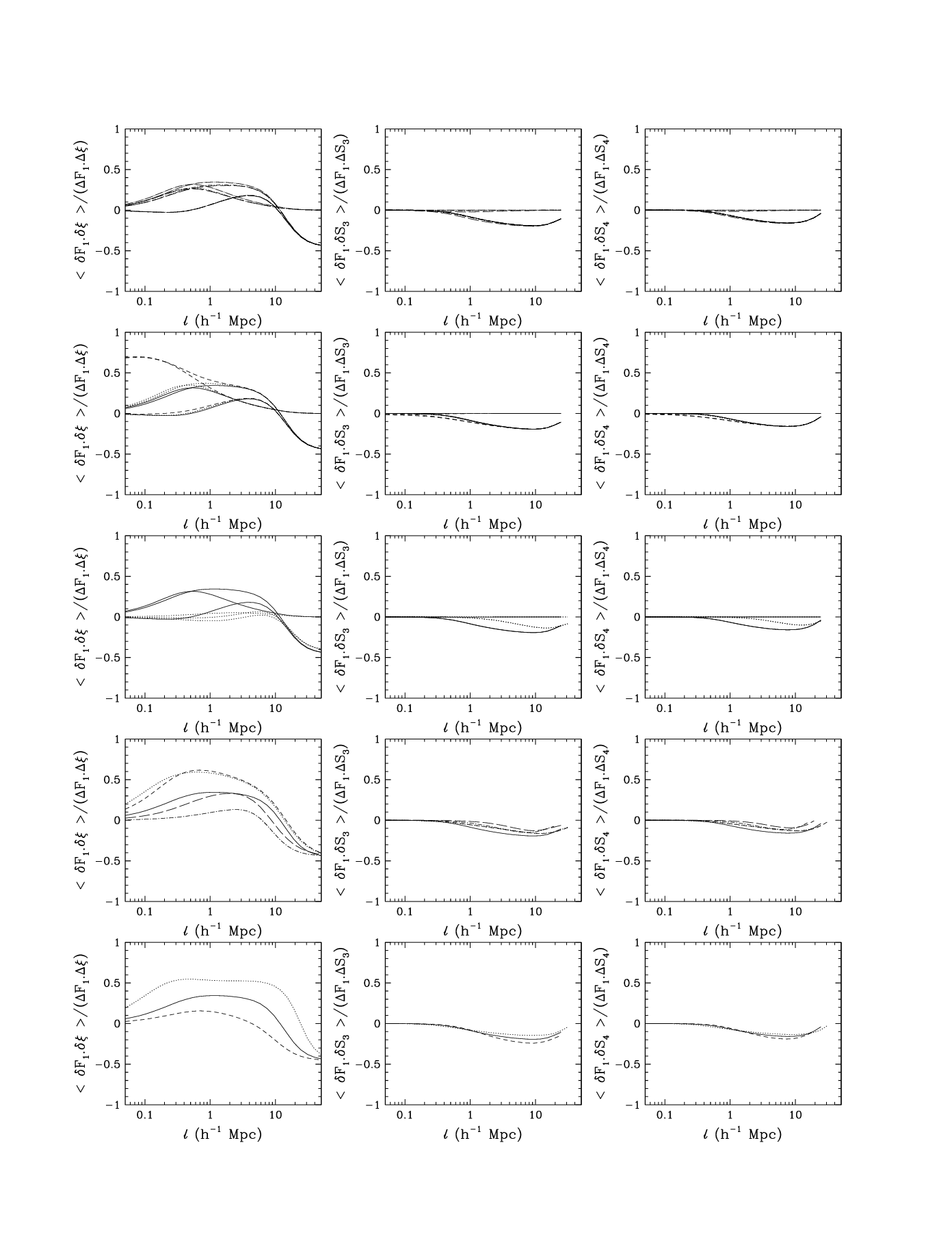,bbllx=11pt,bblly=73pt,bburx=560pt,bbury=800pt,width=13cm}}}
\caption[]{ Same as Figure 5, with the orders $N=1,2,3,4$ corresponding
to the average count $F_1 = \bar N$,  $\xiav$, $S_3=3Q_3$, and $S_4=16
Q_4$, respectively. There are some differences, however, which are listed next.
The range of the $y$ axis is changed to $[-1,1]$. The sequence
of the various contributions in the three upper rows is different
from that of Figure 5, expect for the first column. 
(In the third row of this column the cross-correlations
are approximately zero for the diluted case, and the order
is slightly different but unimportant).
The rest of the columns in the three upper rows have approximately
zero finite volume contributions to the correlation coefficient.
Thus the finite volume effect is  easily identifiable 
as a straight line, while the other
curves all superpose and they correspond to the overlapping and
total contributions.

The right end point of the curves is chosen according to 
equation (\ref{eq:validity}), replacing ``$\ll$'' with ``$\leq$''. 
This condition is not exact, the sharp downturn on 
many panels suggests that a realistic limit is around $10\hmpc$. 
}
\label{fig:figure6}
\end{figure*}
\begin{figure*}
\centerline{\hbox{\psfig{figure=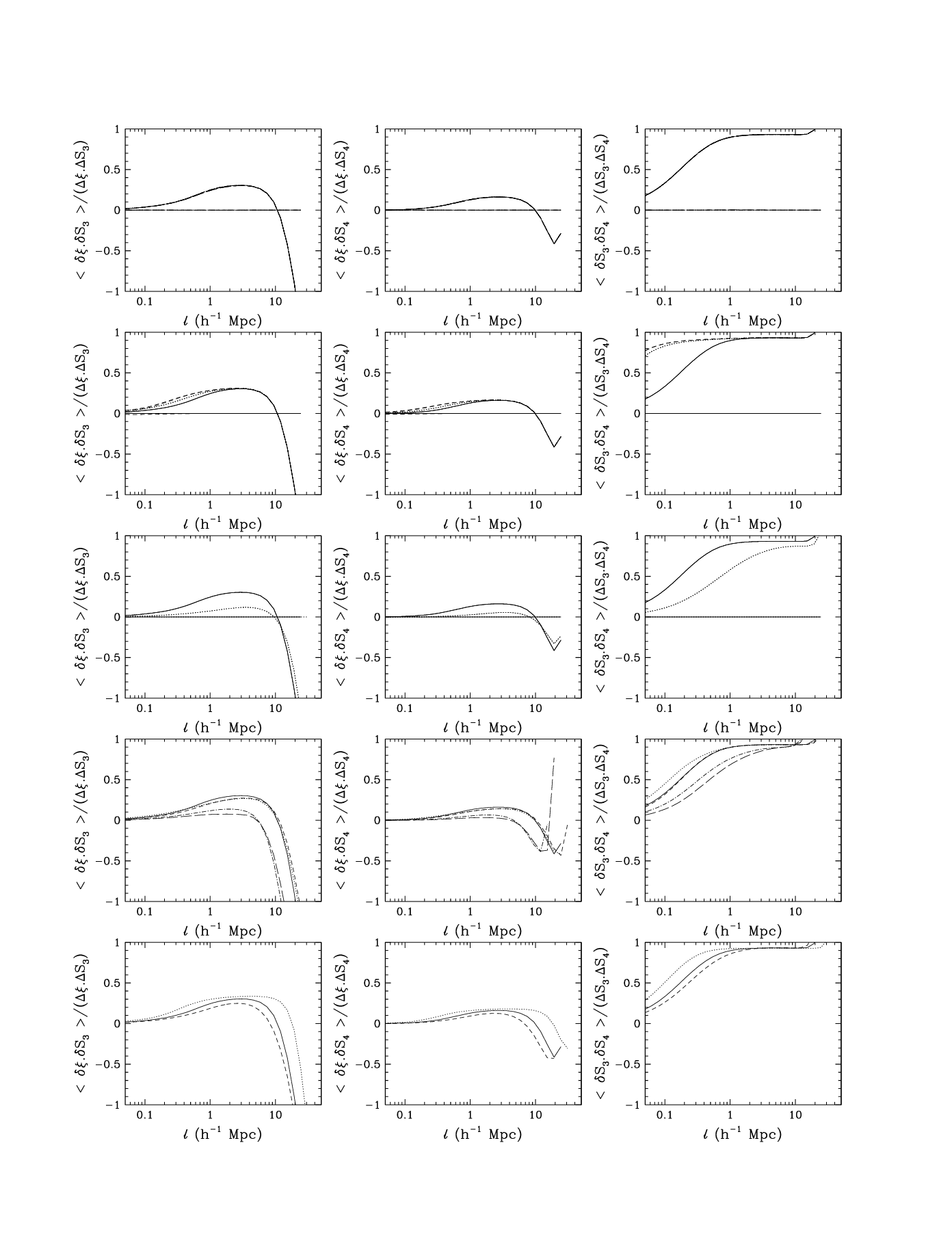,bbllx=11pt,bblly=73pt,bburx=560pt,bbury=800pt,width=13cm}}}
\nonumber
{\small {\bf \noindent Figure 6:} Continued.}
\end{figure*}
Figure 6 displays the correlation coefficients $\delta_{kl}$ for the
cumulants. The figure is exactly analogous to Figure 5.
Similar conclusions can be drawn as previously; we only point out the
differences: i) the perturbative nature of our method limits
the domain of applicability of the results,
ii) finite volume contributions are appreciably weaker 
than for factorial moments, as already established for the cosmic
errors, iii) the dependence on the underlying clustering is
complicated to interpret because of the different ratio natures of the
various statistics involved; this is explained in more detail
below. 

\begin{figure}
\centerline{\hbox{\psfig{figure=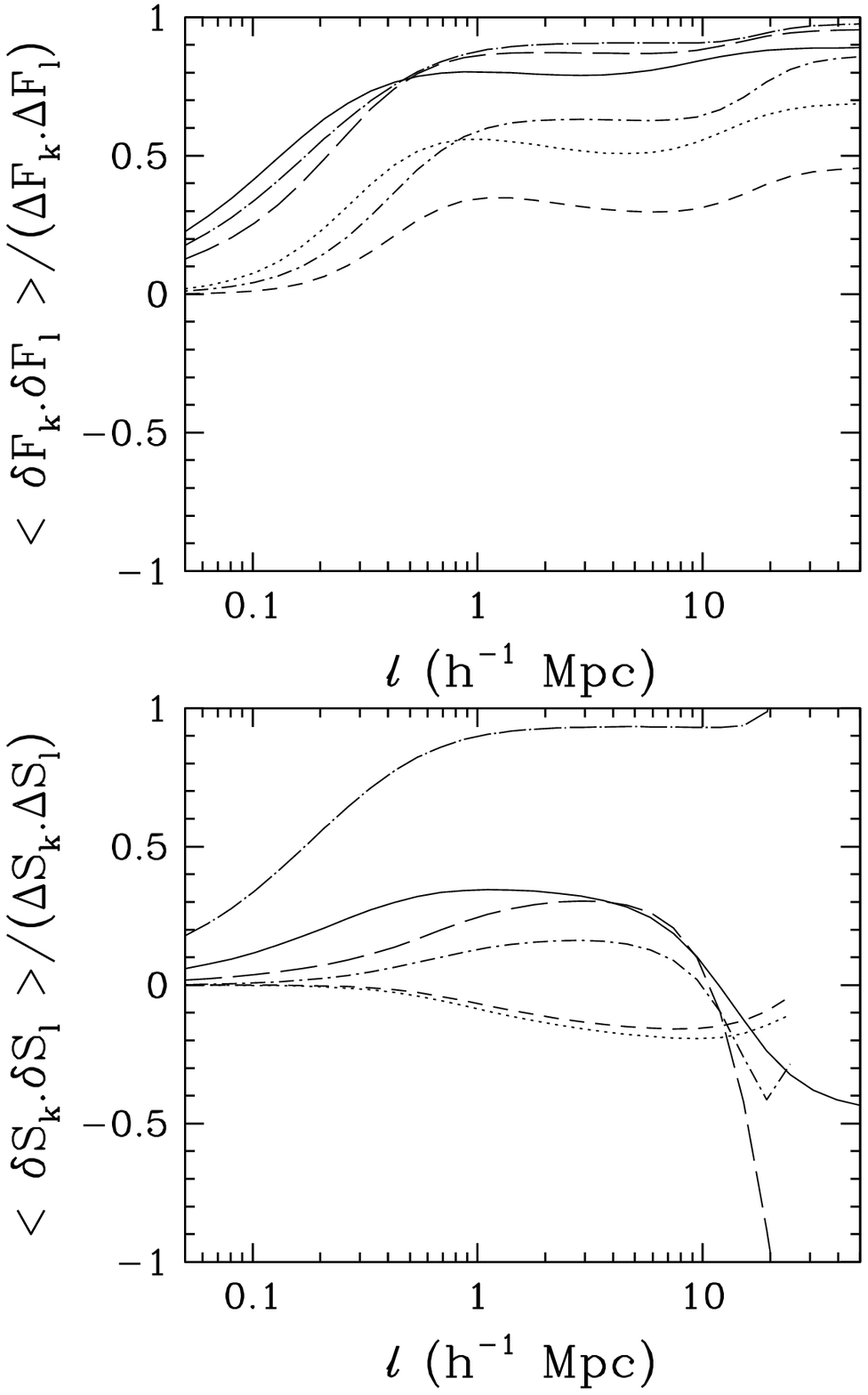,bbllx=137pt,bblly=269pt,bburx=470pt,bbury=800pt,width=8cm}}}
\caption[]{Summary of the cross-correlation results.
The factorial moments (upper panel) and cumulants (lower panel)
are displayed assuming CDM1, E$^2$PT with $n = -2.5$.
The orders $(k,l)$ are distinguished
by different line-types. $(1,2)$: solid,
$(1,2)$:dots, $(1,4)$:dashes, $(2,3)$:long dashes, $(2,4)$:dots-dashes,
$(3,4)$:dots-long dashes. The curves for the lower panel are displayed when 
equation (\ref{eq:validity}) is valid (replacing ``$\ll$'' with ``$\leq$'').
}
\label{fig:figure7}
\end{figure}

Figure 7 illustrates the principal results for
cross-correlations in the SDSS. The factorial moments
are always positively correlated.
The correlations depend on the difference of orders
$|k-l|$, the larger the difference the smaller  the
correlation coefficient, in agreement with intuition. It is worth
noticing that the correlations exhibit approximately the same magnitude
and scale dependence for the same value of $|k-l|$, 
i.e. increase from small scales $\ell \la 1 \hmpc$ 
to a plateau at larger scales. On small scales the correlations
are diluted by discreteness.

The behavior of the cross-correlations for the cumulants is
more difficult to interpret. There are three classes of
cumulants: $\bn$ (order 1), $\xiav$ (order 2), and 
$Q_N$ (order $N$)\footnote{Thus the first
two classes have only one member each.}, 
each with slightly different normalization for
historical and practical reasons: $\xiav$ scales
with $\bn^{-2}$, and the $Q_N$'s likewise with $\xiav^{-(N-1)}$.
Thus one has to interpret separately the correlations
between $\bn$ and $\xiav$,  $\bn$ and $Q_N$'s,  $\xiav$ and $Q_N$'s,
and finally between the $Q_N$'s themselves. The latter are
the simplest to understand: they have similar positive correlations to
the factorial moments, as expected. The rest of the correlations
are fairly weak,  in agreement with intuition when the difference
of orders $|k-l|$ is large. The correlations for $\bn$ and $\xiav$, and
for $\xiav$ and $Q_3$ are smaller than for factorial moments of the
same order. This is due to the ratio nature of $\xiav$ and $Q_3$ which
suppresses the correlations somewhat. As mentioned earlier, 
the perturbative nature of our method limits the validity
of our results above $10\hmpc$ for the SDSS-like surveys.
Also, there are same small negative correlations which should
not be over-interpreted. At the present level of accuracy 
only the weakness of correlations can be established.

\section{Summary and Discussion}

This article formulated the theory of errors on quantities
related to counts in cells, focusing especially 
on cumulants and factorial moments. 
A universal, analytic method based on Taylor expansion approach was devised
to calculate \underline{\bf explicitly} the {\em cosmic error}, 
the {\em cosmic bias}, 
and the  {\em cosmic cross-correlations} for virtually any statistics
derived from counts in cells. There are always three contributions
to these quantities \cite{sc96}: finite volume, edge, and discreteness
effects. The principal
results are the following:

\begin{enumerate}

\item {\em Cosmic errors:}
SC have computed the cosmic errors on factorial moments
for two particular cases of the hierarchical model. CSS
have extended the results for inhomogeneous catalogs and for optimal
weighting. These previous calculations 
have been generalized for cumulants, and for 
PT; explicit analytic results for the factorial moments are given
in Appendix A. The cosmic error depends on the
bivariate distributions, for which EPT had to be generalized.
The new Ansatz is termed E$^2$PT, and explained in detail
in Colombi et al.~(1999b). 
For the SDSS it is predicted that
the cumulants fare better than the factorial moments
on scales $\ell \la 10\hmpc$.  On large scales the situation is
reversed  due to the enhanced sensitivity of the connected 
moments to edge effects. For the particular example of the SDSS, however, 
this regime is outside the validity of our perturbative method.
In the scale range of $1\hmpc - 10\hmpc$
the expected errors are smaller than 3 \% for $\xiav$,
4 \% for $Q_3$, and 15 \% for $Q_4$. For reference, 
the errors determined by CSS for factorial moments
of order $k = 2,3$, and $4$ were 
$1-2\%$, $3-5\%$, and, $5-10\%$, respectively, in the
regime  $1\hmpc \la \ell \la 50\hmpc$.
A detailed investigation in
a range of reasonable models shows that the estimates
are robust within a factor of $\sim 2$. 

Note that according to equation (8) $\xiav$ is a linear
functional of $\xi$, the two point correlation function.
In fact, if $\xi$ is a power-law of index $\gamma$
the two are proportional to each other
$\xiav \propto \xi$. For a linear functional, the error
propagation is expected to be especially simple: 
the errors on $\xi$ should be a linear function 
of the errors on $\xiav$. Of course, this statement is
only approximate, because its validity depends on the nature of
the estimators used to measure $\xi$ and $\xiav$.
For a power-law correlation function,
we conjecture that the approximation 
\begin{equation}
	\sigma_{\xi} \simeq  \sigma_{\xiav}
\label{eq:sigxi}
\end{equation}
holds for the relative cosmic error. There
might be some difference at large scales, where edge effect
dominate and can be at least partly corrected for estimators
of $\xi$ (e.g., Ripley 1988; Landy \& Szalay 1993; Szapudi \& 
Szalay 1998)  but not for standard estimators of $\xiav$ (e.g.,
CSS). In that regime, it is therefore expected that
$\sigma_{\xi} \la \sigma_{\xiav}$.
Nevertheless, approximation (\ref{eq:sigxi}) 
should be more accurate for estimating
the errors on the two point correlation function than the 
methods prevailing in the literature, especially the 
meaningless bootstrap method.

\item {\em Cosmic bias:} 
an estimator is biased if its
ensemble average is different from the true value.
This is typical when non-linear
functions of unbiased estimators are constructed, 
such as $\xiav$, and the $Q_N$'s. 
For such statistics a perturbative expansion can be used to
determine the bias $b$. A simple but important consequence is
that $b  = {\cal O} (\sigma^2)$, where $\sigma$ is the relative
cosmic error. As a result the cosmic bias
is negligible compared to the cosmic error in the perturbative
regime. A necessary, and in practice sufficient \cite{hubble1}, 
 criterion for the 
validity of the series expansion is that $b \ll \sigma \ll 1$.
For the SDSS the cosmic bias is predicted to be negligible
on scales $\la 50\hmpc$ for $\xiav$, and $\la 10\hmpc$ for
higher order statistics. Explicit formulae are given for
$b_{\xiav}$ and $b_{Q_3}$ in Appendix B. 

\item {\em Cosmic cross-correlations:} 
they generalize the concept of the cosmic
error by considering the full correlation matrix of the 
statistics.
Correlations between indicators influence the constraining 
power of measurements on theories.
The calculation for the cross-correlations
of the factorial moments is  exactly analogous to that of
the cosmic error presented by SC. {\em Explicit
analytic} results are given in Appendix A. Together
with the results of SC this completes the
theory of the full cosmic cross-correlation matrix and
forms the basis of subsequent calculations concerning
the errors of any quantity related to counts in cells,
such as the cumulants. 

While the following results were established in a concrete
example, i.e. a suit of SDSS like surveys, we conjecture that they
are quite generic.
In agreement with intuition factorial moments of close orders
appear to exhibit stronger correlations than those of far orders.
The results are more complex for cumulants, although the $Q_N$'s
behave similarly to factorial moments. Interestingly,
the correlations between $\xiav$ and $\bn$, and between $\xiav$
and $Q_N$'s are weaker than for factorial moments of the same
order. Optimal weighting naturally augments correlations, and
discreteness effects likewise reduce them. These results
depend significantly on the clustering properties of the
underlying distribution of galaxies, although the qualitative
features are robust. 

\end{enumerate}

The theoretical calculations of this paper were confronted
with measurements in a state of the art large $\tau$CDM
simulation \cite{hubble1}; the
results are previewed next.

The detailed investigations
suggest that the theory of errors presented in this article
is fairly accurate, especially in the weakly non-linear
regime, where a few percent precision was achieved for
the factorial moments. In the
highly non-linear regime it appears that the approximate
nature of the models for bivariate distribution translates
into a slight overestimation of the errors, perhaps by a 
factor of two in the worst case. The situation will be improved
in the future, if more realistic models are constructed
for the bivariate counts. 

The predicted cross-correlations for the factorial moments
describe the qualitative
features of the measurements quite well, however, the details
are less precise than for the errors. When the difference
of orders $|k-l| = 1$, the theory
is  about $20\%$ accurate, while it gradually looses
precision, up to about $50\%$ in the worst case,  
as the difference of orders increases. This behavior suggests
that the underlying locally Poisson assumption becomes
less precise.
An attempt to improve on this would
introduce encumbering complications 
because of the necessity of the trivariate
generating function, and is left for future research.

The present results complement the investigations
of SC, and their generalization 
by CSS for inhomogeneous catalogs.
Together they constitute the statistically complete
description of the errors whenever
the Gaussian approximation for the cosmic distribution 
of events is sufficiently accurate. 
This is true  when the cosmic errors are small
\cite{hubble2},  an essential result for likelihood analyses. 
Applications of the theory of cross-correlations are
discussed elsewhere \cite{scb99b,bcs299}. 

While the investigations presented in this article are
sufficiently accurate for any foreseeable practical application and
included all crucial effects and contributions, there are
some minor points which were not mentioned thus far:

\begin{enumerate}

\item{\em Galaxy bias} (not to be confused with the cosmic 
bias): light might not trace mass, thus the
statistical properties of galaxies might be different 
from those of the dark matter. Theories and models relying
only on dark matter dynamics such as PT and EPT might
miss some important aspects of the galaxy distribution.
However, current measurements in two and three dimensional
galaxy catalogs  suggest that the models used here
such as SS, BeS, and even EPT, yield fairly realistic
description (e.g., Gazta\~naga 1994; Szapudi, Meiksin \& Nichol
1996). To be complete, however, one should
in principle include the effects of bias in the theory.

\item {\em Redshift distortions:} they
arise from the peculiar velocities of galaxies in three dimensional
catalogs. Their effect on the statistics is well known.
The two-point correlation function
and the amplitude of the $Q_N$'s decreases in the highly non-linear
regime, while in the weakly non-linear regime only the normalization
of the two-point correlation function is affected significantly
(e.g., Matsubara \& Suto 1994; Hivon et al.~1995; Szapudi et al.~1999b).
The extent to which redshift
distortions alter clustering is thus well within the range of 
variations considered previously.

\item {\em Cosmological parameters :}
the dependence of the $Q_{N}$ coefficients on cosmological parameters 
is extremely weak. This has been explicitly shown in PT
\cite{Bouchetetal92,bern94a,hivon}, and  it is expected to carry over to
the nonlinear regime as well (Nusser \& Colberg 1998; 
Scoccimarro \& Frieman 1998; Szapudi at al.~1999b).

\item{\em Angular catalogs and weak lensing:}
this article considered three dimensional distributions only. 
Analogous calculations can be done for angular catalogs,
and for weak lensing which promises to be an important mean
of investigation of the cosmological parameters in the near
future \cite{BvWM97,JSW99,GB98}. This point is investigated
elsewhere \cite{bcs99}.

\item {\em Edge effects:} so far the calculations were
performed to leading order in $v/V$, and the results are independent
of the geometry of the catalog. This is sufficiently precise
approximation for compact surveys such as the SDSS. However,
for more complicated survey geometries, such as the 2dF
or the VIRMOS survey, the computations can be improved by
taking into account higher order terms. The next to leading
order term depends on the perimeter (surface) of the survey.
This is studied in Colombi et al.~(1999).

\item {\em Full description:} for maximum likelihood analyses
with multi-scale measurements there is one more step needed
to complete the statistical description. The cross-correlation
matrix should be calculated between statistics estimated
on different scales. This calculation is a trivial although
somewhat tedious generalization of the previous considerations.
It is left for future work.

\item {\em Cosmic bias:}
these results are in contrast with that of Hui \& Gazta\~naga (1998,
HG). The reasons
for the difference are that a) they neglected discreteness effects,
which could be significant on small scales for cumulants $Q_N$, b)
although their calculation in principle includes edge effects
dominant on large scales, they finally neglected them (however, see
the discussion in Appendix B), c) they did not realize
that  $b  = {\cal O} (\sigma^2)$ in the perturbative regime.
Outside of the domain of validity, this condition naturally
breaks as the measurement of HG suggests. However,
to estimate the cosmic bias and the cosmic error they use
only 10 realizations of the local universe. In the Virgo
Hubble Volume simulation with 4096 realizations, 
Colombi et al.~(1999b)
find that the cosmic bias is always dominated by the 
cosmic errors. Moreover, according to 
Szapudi et al.~(1999c),
the cosmic distribution function, the probability
distribution function of measurements, shows significant skewness.
This is a source of effective bias for only one
realization, i.e. our local universe; see Colombi et al.~(1999b)
and Szapudi et al.~(1999c) for a detailed discussion.
HG have proposed an Ansatz for scales
beyond the validity of Taylor expansion in the theory. 
This recipe, however, neglects edge effects, 
which constitute the dominant contribution
on large scales, except for $\xiav$ (see Appendix B); 
the apparent agreement of their Ansatz
with measurements appears to be a coincidence.
Nevertheless, their calculations, if
sufficiently tested and gauged with $N$-body
experiments, may be still used to estimate the cosmic bias.
A detailed comparison of their analytic results for $\xiav$ with ours is 
contained in the Appendix B.
\end{enumerate}

\section*{Acknowledgments} 

I.S. was supported by the PPARC rolling grant for 
Extragalactic Astronomy and Cosmology at Durham.
F.B. thanks IAP, I.S. IAP and Saclay, S.C. University of Durham 
for their hospitality.


\def\apj {ApJ}
\def\aap {A \& A}
\def\ajs{ApJS}
\def\apjs{ApJS}
\def\mnras {MNRAS}

\appendix
\onecolumn
\section{The Cosmic Error and Cross-Correlations for Factorial Moments}
This section complements the analytic results for the cosmic
errors obtained in SC with explicit formulae for the cross-correlations.
These together with the previous results establish the full cosmic
cross-correlation matrix, which underlies all error calculations
for statistics related to counts in cells.

For the sake of conciseness and simplicity the following notation
is introduced for the cosmic cross-correlation matrix
\begin{equation}
  \Delta_{kl}\equiv {\rm Cov}(F_k,F_l)=\langle \delta F_k \delta F_l \rangle.
\end{equation}
Note that $\Delta_{kk}=(\Delta F_k)^2$ is the cosmic error.
$\Delta_{kl}$ has three contributions
\begin{equation}
   \Delta_{kl}=\Delta_{kl}^{\rm F}+\Delta_{kl}^{\rm E}+\Delta_{kl}^{\rm D},
\end{equation}
where $\Delta_{kl}^{\rm F}$, $\Delta_{kl}^{\rm E}$ and 
$\Delta_{kl}^{\rm D}$
are the finite volume, edge and discreteness effect contributions,
respectively.
SC computed $\Delta_{kk}^{\rm F}$, $k \leq 4$, and presented
the analytic results for $k \leq 3$,
within the framework of the SS and the BeS models.
Assuming local Poissonian behavior, and 
a power-law  $r^{-\gamma}$ for $\xi(r)$ on
scales $r\leq 2\ell$, they also calculated 
the discreteness and edge effect contributions, $\Delta_{kk}^{\rm D}$
and $\Delta_{kk}^{\rm E}$ for $k \leq 4$ with explicit formulation
for $k \leq 3$.  All computations were performed to leading order in $v/V$, 
where $v$ and $V$ are the cell and the sample
volume, respectively. The aim of this Appendix is to 
present the extension of their
results for PT (and EPT) for the finite volume
contribution (Appendix A.1), and for
cross-correlations $k < l \leq 3$
(Appendix A.2). Note that, as in SC, all the calculations were performed
up to fourth, but the results are only printed to third order.
A FORTRAN program can be obtained from the authors for computing
numerically the cosmic errors, cross-correlations and biases for
factorial moments and cumulants.

\subsection{The Finite Volume Error 
for Factorial Moments in PT and E$^2$PT framework}

The bivariate generating function
for counts in cells employed in SC had to be generalized to incorporate PT.
This generalization can be used for most other models,
including SS and BeS. The explicit results 
from this formalism are presented next:
\begin{equation}
 \Delta_{11}^{\rm F}={{\bn}^2}\,\xiav({\hat L}),
\end{equation}
\begin{equation}
 \Delta_{22}^{\rm F}=4 {{\bn}^4}\,\xiav({\hat L})\,
  \left( 1 + 2\,\xiav\,Q_{12} + 
    {{\xiav}^2}\,Q_{22} \right),
\end{equation}
\begin{equation}
\Delta_{33}^{\rm F}=9{{\bn}^6}\,\xiav({\hat L})\,
  \left( 1 + 2\,\xiav + {{\xiav}^2} + 
    4\,\xiav\,Q_{12} + 
    4\,{{\xiav}^2}\,Q_{12} + 
    6\,{{\xiav}^2}\,Q_{13} + 
    6\,{{\xiav}^3}\,Q_{13} + 
   4\,{{\xiav}^2}\,Q_{22} + 
    12\,{{\xiav}^3}\,Q_{23} + 
    9\,{{\xiav}^4}\,Q_{33} \right).
\end{equation}
The quantity $\xiav({\hat L})$ is roughly
the average of the two-point correlation function over the
survey volume:
\begin{equation}
 \xiav({\hat L}) \equiv \frac{1}{{\hat V}^2} \int_{|r_1-r_2| \geq 2\ell}
   \rmd^3r_1 \rmd^3r_2 \xi(|r_1-r_2|).
\end{equation}
To leading order in $v/V$ this integral reads (Colombi et al.~1999a)
\begin{equation}
  \xiav({\hat L}) \simeq \xiav_0({\hat L})
  - \frac{8v}{\hat V} {\xiav}_1(2\ell),
  \label{eq:xilbox}
\end{equation}
with
\begin{equation}
 \xiav_0({\hat L})=\frac{1}{V^2} \int_{\hat V} \rmd^3r_1 \rmd^3r_2 \xi(|r_1-r_2|),
\end{equation}
\begin{equation}
 \xiav_1(\ell)=\frac{1}{v}\int_{r \leq \ell} 4\pi r^2 \rmd r \xi(r).
\end{equation}
For most practical cases, the term proportional to $\xiav_1(2\ell)$ 
can be neglected and the integral can be performed
on the sample volume $V$ instead of the volume covered
by the cells included in the catalog, ${\hat V}$: $\xiav({\hat
L})\simeq \xiav_0(L)$.
If kept, the correction $8v \xiav_1(2\ell)/{\hat V}$, 
which can be viewed as an ``edge-finite volume
effect'', yields usually a small
correction compared to the edge effect errors (see Colombi et al.~1999a,b for
practical examples). 

In the PT framework, the cumulants factorize 
$Q_{kl}=Q_{k1}Q_{l1}$. Each $Q_{k1}$ depends on 
logarithmic derivatives $\gamma_j=-n_j-3$ of the
(linear) variance, $\xiav$, with respect to scale \cite{bern96}.
Note that in the E$^2$PT framework, the nonlinear variance $\xiav$ is taken.
The parameter $\gamma_1$ is adjusted such 
that $S_3=3 Q_3=34/7+\gamma_1$ fits the measured, 
nonlinear skewness. 
Higher order statistics and bivariate statistics are then derived from
PT expressions with this value of 
$\gamma_1$ (and $\gamma_j=0$, $j \geq 2$). A detailed numerical 
investigation of E$^2$PT for the cosmic errors can be 
found elsewhere \cite{hubble1}.

The above results can represent the SS model as well 
by replacing $Q_{kl}$ with
$Q_{k+l}$. In the BeS framework, similarly as in PT, 
the relation $Q_{kl}=Q_{k1}Q_{l1}$ holds. In that case 
the $Q_{k1}$ can be computed explicitly from
the vertex generating function as combinations
of $Q_l$, $\ell \leq k+1$ (See BeS and SC for
details). Corresponding 
analytic expressions of the finite volume error 
can be found in SC.

Note finally that for the BeS and PT models, because of the factorization
properties (\ref{eq:Qfact}) and (\ref{eq:QfactPT}), we have
\begin{equation}
{\Delta^{\rm F}_{kl}\over{\bn}^{k+l}\xiav(L)}=
{\Delta^{\rm F}_{k1}\over{\bn}^{k+1}\xiav(L)}\ 
{\Delta^{\rm F}_{l1}\over{\bn}^{l+1}\xiav(L)}.
\end{equation}

\subsection{The Cosmic Cross-Correlations for Factorial Moments}
The explicit formulae of the cosmic cross-correlations presented
next complete the cosmic cross-correlation matrix. They
provide the full statistical description to second order
and can be used both for maximum likelihood analysis, and for
calculating the cross-correlation matrix of any estimator related
to factorial moments with the method presented in the main text.
\begin{equation}
 \Delta_{12}^{\rm F}=2{{\bn}^3}\,\xiav({\hat L})\,
  \left( 1 + \xiav\,Q_{12}
    \right),
\end{equation}
\begin{equation}
\Delta_{12}^{\rm E}={{\bn}^3} \xiav \frac{v}{V}
  \left( 8.525+ 
  11.42\,{\xiav}\,Q_3 \right),
\end{equation}
\begin{equation}
\Delta_{12}^{\rm D}={{\bn}^2} \frac{v}{V} \left(2.0 + 
  1.478\,\xiav \right),
\end{equation}
\begin{equation}
 \Delta_{13}^{\rm F}=3 {{\bn}^4}\,\xiav({\hat L})\,
  \left( 1 + \xiav + 
    2\,\xiav\,Q_{12} + 
    3\,{{\xiav}^2}\,Q_{13} \right),
\end{equation}
\begin{equation}
\Delta_{13}^{\rm E}={{\bn}^4} \xiav \frac{v}{V}
  \left( 9.05 + 
  11.42\,{{\xiav}} + 
  21.67\,{{\xiav}}\,Q_3 + 
  42.24\,{{\xiav}^2}\,Q_4 \right),
\end{equation}
\begin{equation}
\Delta_{13}^{\rm D}={{\bn}^3} \frac{v}{V}
  \left( 3.0 + 
  6.653\,\xiav + 
  4.949\,{{\xiav}^2}\,Q_3 \right),
\end{equation}
\begin{equation}
 \Delta_{23}^{\rm F}=6 {{\bn}^5}\,\xiav({\hat L})\,
  \left( 1 + \xiav + 
    3\,\xiav\,Q_{12} + 
    3\,{{\xiav}^2}\,Q_{13} + 
    {{\xiav}^2}\,Q_{12} + 
    2\,{{\xiav}^2}\,Q_{22} + 
    3\,{{\xiav}^3}\,Q_{23} \right),
\end{equation}
\begin{equation}
 \Delta_{23}^{\rm E}={{\bn}^5} \xiav \frac{v}{V}
  \left( 23.08 + 
  33.09\,{{\xiav}} + 
  90.17\,{{\xiav}}\,Q_3 + 
  55.19\,{{\xiav}^2}\,Q_3 + 
  211.2\,{{\xiav}^2}\,Q_4 + 
  229.9\,{{\xiav}^3}\,Q_5
  \right),
\end{equation}
\begin{equation}
\Delta_{23}^{\rm D}={{\bn}^3}\frac{v}{V}
 \left( 1.943 + 6.\,{{\bn}} + 
  4.522\,\xiav + 
  26.61\,{{\bn}}\,\xiav + 
  9.898\,{{\bn}}\,{{\xiav}^2} + 
  3.531\,{{\xiav}^2}\,Q_3 + 
  39.59\,{{\bn}}\,{{\xiav}^2}\,Q_3 + 
  39.53\,{{\bn}}\,{{\xiav}^3}\,Q_4 
 \right).
\end{equation}
In the above equations the edge and discreteness effect contribution 
was calculated from a locally Poisson Ansatz.
On scales smaller than twice the cell size
the two-point correlation function is assumed  to be a power 
law $\xi(r) \propto r^{-\gamma}$
with $\gamma=1.8$. 
Detailed investigation of SC showed that variations of $\gamma$ affect
insignificantly the coefficients in the above equations. Therefore
these equations are valid even
when $\xi$ departs weakly from a strict power-law.

\section{The Cosmic Bias: Comparison with HG}
\subsection{The cosmic bias on $\xiav$: detailed analysis}
Within the theoretical framework of this article, 
the cosmic bias on $\xiav$ can be expressed
in terms of $\Delta_{kl}$ (defined in Appendix~A):
\begin{equation}
  b_{\xiav}=\frac{F_2}{\xiav F_1^2} 
\left( 3\delta_{11} - 2 \delta_{12}\right).
\end{equation}
with
\begin{equation}
  \delta_{kl}\equiv \frac{\Delta_{kl}}{F_k F_l}.
\end{equation}
Using the analytic results in Appendix A and
assuming E$^2$PT,
the cosmic bias can be written to leading order in $v/V$ as
\begin{equation}
  b_{\xiav}=b_{\rm D} + b_{\rm E} + b_{\rm F},
\end{equation}
where the discreteness, edge, and finite volume effects are,
respectively,
\begin{equation}
  b_{\rm D}\simeq \left( -\frac{1}{\xiav} + 0.04 \right) \frac{v}{\bn V},
  \label{eq:disbia}
\end{equation}
\begin{equation}
  b_{\rm E}\simeq \left (-\frac{16.5}{\xiav} + 16.5 - 18.5 Q_3 \right) \frac{\xiav v}{V},
  \label{eq:bedge}
\end{equation}
\begin{equation}
  b_{\rm F} \simeq \left( -\frac{1}{\xiav} + 3 - 2 Q_{12} \right)
  \xiav({\hat L}).
  \label{eq:bfinite}
\end{equation} 
The result of HG is the following
\begin{equation}
  b_{\xiav}= \left( -\frac{1}{\xiav} + 3 - 2 Q_{12} \right) \xiav_2^L,
\end{equation}
with
\begin{equation}
 \xiav_2^L\equiv \frac{1}{{\hat V}^2} \int \rmd^3r_1 \rmd^3r_2 
{\breve \xi}(|r_1-r_2|),
 \label{eq:xiavHG}
\end{equation}
\begin{equation}
 {\breve\xi}(r)=\int_{v_1,v_2} \rmd^3x_1 \rmd^3x_2 \xi(|x_1-x_2|). 
\end{equation}
The integral in the above equation is performed over two cells 
with volumes $v_1,v_2$ separated
by distance $r$. 
Thus the calculation of HG drops discreteness effects, claiming
that they can be neglected since
$v/(\bn V)=1/N_L$ is small. In contrast, SC have
shown that terms proportional to $1/N_L$ are dominating the
cosmic error on small scales. This may be true in principle 
for the cosmic bias as well. Equation (\ref{eq:disbia}), however, shows
that discreteness effects are indeed negligible,
unless $\xiav \sim 1/N_L$. Note that the same argument is invalid for
higher order cumulants  such as $Q_3$ and $Q_4$: there
discreteness effects can induce a {\em significant} 
contribution to bias, particularly
on small scales (see the example below). 

The calculation of HG includes edge effects
through the integral (\ref{eq:xiavHG}) over the volume
${\hat V}$ covered by cells {\em included} in the catalog. 
Following SC one can split integral (\ref{eq:xiavHG}) 
into two contributions according to whether the cells overlap or not
\begin{equation}
   \xiav_2^L=\frac{1}{{\hat V}^2} \left[ \int_{|r_1-r_2| \geq 2\ell} \cdots
+\int_{|r_1 - r_2 | \leq 2\ell} \cdots \right].
\end{equation}
While it would be superfluous here to enter into details of this
somewhat tedious calculation, it is clear, 
as in SC, that the overlapping term will
typically yield a contribution $b_{\rm E}$ proportional to $\xiav v/V$. 
On the other hand, disjoint cells contribute approximately  
of order $\xiav({\hat L})=\xiav_0({\hat L})-8(v/{\hat V})
\xiav_1(2\ell)$. [This reasoning is valid to leading order 
in $v/V$. Higher order corrections proportional 
to the perimeter of the survey must be
taken into account for more accuracy \cite{virmos}]. 
Since the correction proportional to $\xiav_1$ might exactly
compensate for the term $b_{\rm E}$ introduced by overlapping
cells, HG argue that $\xiav_2^L \simeq \xiav_0(L)$,
suggesting exact cancellation.
Our calculations based on local Poisson approximation indeed  
show that $b_{\rm F}/\xiav({\hat L})$ is of same order of
$b_{\rm E}/[(8v/{\hat V}) \xiav_1(2\ell)]$ for the
particular case of $\xiav$. This result does
not hold, however, for cumulants of higher order, where edge
effects are dominant on large scales.
At this level of accuracy our calculation becomes approximate
as well mainly because of the local Poisson assumption
(Colombi et al.~1999b), therefore it is impossible to
evaluate the residual edge effects for $\xiav$ in this
framework.


\subsection{The cosmic bias on higher order statistics}

A simple algebraic calculation of
the cosmic bias on $Q_3=S_3/3$ yields
\begin{equation}
  b_{Q_3}=b_{\xiav_3}-3 b_{\xiav}-2 \delta_{23} + 3 \delta_{22},
 \label{eq:bs3}
\end{equation}
with
\begin{equation}
  b_{\xiav_3}=\frac{F_3}{\xiav_3 F_1^3} (6 \delta_{11} - 3 \delta_{13})
  -3 \frac{F_2}{\xiav_3 F_1^2} (3 \delta_{11} - 2 \delta_{12}).
\end{equation}
Explicit writing of the discreteness 
contribution in equation (\ref{eq:bs3}), although
trivial, would go beyond the scope of this paper. 
To illustrate that it is not negligible,
numerical results are given next. For
$\ell=1\hmpc$, $b_{\xiav}\simeq -5\times 10^{-5}$, 
$b_{Q_3}=-2\times10^{-4}$ 
in the standard SDSS-like catalog of CSS. After
a dilution by a factor 100 (which means that the catalog would
still contain $\sim 8000$ objects, e.g.~CSS), 
these terms become $b_{\xiav}=-4\times 10^{-5}$, a small change
as expected, and
$b_{Q_3}=-0.2$, a change by three orders of magnitude.
This means that discreteness effects can have a significant
contribution to the bias on small scales, 
in contrast with the claims of HG. The accuracy
of this statement is limited by
the local Poisson assumption, which is, however,  increasingly
more precise as the the sample becomes more and more
diluted.
\end{document}